\documentclass[preprint,prd,aps,showkeys,nofootinbib]{revtex4}
\usepackage{graphicx}
\usepackage{color}
\usepackage{CJK,upgreek,fancyhdr}
\usepackage{bm}

\begin{document}

\begin{CJK*}{GBK}{song}

%\preprint{hep-ph/0412147}

\title{Explaining the possible 95 GeV excesses in the B-L symmetric SSM}

\author{Wen-Hui Zhang$^{1,2}$\footnote{zwh\_0218@163.com },Jin-Lei Yang$^{1,2,3}$\footnote{jlyang@hbu.edu.cn},Ming-Hui Guo$^{1,2}$,Hai-Bin Zhang$^{1,2,3}$,Tai-Fu Feng$^{1,2,3}$}

\affiliation{Department of Physics, Hebei University, Baoding, 071002, China$^1$\\
Key Laboratory of High-precision Computation and Application of Quantum Field Theory of Hebei Province, Baoding, 071002, China$^2$\\
Research Center for Computational Physics of Hebei Province, Baoding, 071002, China$^3$}

\begin{abstract}
This study investigates the excesses observed in the diphoton and $b\bar b$ data around $95\;{\rm GeV}$ within the framework of the $B-L$ supersymmetric model (B-LSSM). Comparing with the minimal supersymmetric standard model, the B-LSSM incorporates two singlet chiral Higgs bosons which mix with the SM-like Higgs boson due to the gauge kinetic mixing effect. The richer Higgs sector indicates that designating the B-LSSM specific CP-even Higgs state as the lightest Higgs boson has great potential to explain the excesses at around $95\;{\rm GeV}$. Considering the two-loop effective potential corrections, it is found that the B-LSSM is hard to fit the $125\;{\rm GeV}$ Higgs signal strengthes and the two excesses around $95\;{\rm GeV}$ in the experimental $1\sigma$ intervals, while the model can reproduce them in the experimental $2\sigma$ intervals simultaneously. And considering the two-loop effective potential corrections to the squared Higgs mass matrix is important to account for the mixing effects among Higgs sector.
\end{abstract}


\maketitle

\section{Introduction\label{sec1}}

The discovery of a $125\;{\rm GeV}$ Higgs boson at the Large Hadron Collider (LHC) in 2012~\cite{CMS:2012qbp,ATLAS:2012yve} is one of the most remarkable achievements in theoretical physics, and its properties are consistent well with the standard model (SM) predictions~\cite{CMS:2022dwd,ATLAS:2022vkf} which indicates all of the fundamental particles predicted by the SM are observed. Then identifying whether the detected Higgs boson is the only fundamental scalar particle or part of a new physics (NP) theory with extended Higgs sectors becomes the prime goal of the current LHC programme. Although there is no new scalar was found at the LHC so far, several intriguing excesses in the searches for light Higgs bosons below 125 GeV have been observed with increasing precision in the measurements of Higgs couplings to fermions and gauge bosons.

The results based on both the CMS Run 1 and the first year of CMS Run 2 data for Higgs boson searches in the diphoton final state show a $2.8 \sigma$ local excess at the mass about $95\;{\rm GeV}$~\cite{CMS:2015ocq,CMS:2018cyk}, which is compatible with the latest ATLAS result~\cite{ATLAS:2023CA} based on the previously reported result utilizing $80\;{\rm fb}^{-1}$~\cite{ATLAS:2018xad} in the diphoton searches
\begin{eqnarray}
&&\mu(\Phi_{95})_{\gamma\gamma}=\frac{\sigma(gg\rightarrow \Phi^{\rm NP}_{95}){\rm BR}(\Phi^{\rm NP}_{95}\rightarrow \gamma\gamma)}{\sigma(gg\rightarrow h^{\rm SM}_{95}){\rm BR}(h^{\rm SM}_{95}\rightarrow \gamma\gamma)}=0.18\pm0.10.
\end{eqnarray}
Utilizing the full Run 2 data set, CMS published the results for the Higgs boson searches in the $\tau^+\tau^-$ channel which show a
local significance of $3.1\sigma$ at a mass value about $95\;{\rm GeV}$~\cite{CMS:2022goy}
\begin{eqnarray}
&&\mu(\Phi_{95})_{\tau\tau}=\frac{\sigma(gg\rightarrow \Phi^{\rm NP}_{95}){\rm BR}(\Phi^{\rm NP}_{95}\rightarrow \tau\tau)}{\sigma(gg\rightarrow h^{\rm SM}_{95}){\rm BR}(h^{\rm SM}_{95}\rightarrow \tau\tau)}=1.2\pm0.5.\label{eq2}
\end{eqnarray}
The analysis on the Higgs boson searches in the diphoton final state by CMS further confirmed the excess at about $95\;{\rm GeV}$, and the result reads~\cite{CMS:2023yay}
\begin{eqnarray}
&&\mu(\Phi_{95})_{\gamma\gamma}=0.33_{-0.12}^{+0.19}.\label{eq3}
\end{eqnarray}
In addition, the results measured at the Large Electron Positron (LEP) show a local excess of $2.3\sigma$ at the invariant mass of $b\bar b$ around 98 GeV\footnote{The excess at LEP is broad due to the hadronic $b\bar b$ final state, so this result is also compatible with a scalar resonance at 95 GeV, consistent with the diphoton and ditau excesses. We also take the excess for $b\bar b$ final state at about 95 GeV in the following analysis.}~\cite{LEP:2003ing}, which can be expressed in terms of a signal strength as
\begin{eqnarray}
&&\mu(\Phi_{95})_{b\bar b}=0.117\pm0.057.\label{eq4}
\end{eqnarray}

Theoretically, there are numerous discussions on the excesses in NP models. The analysis carried out in Refs.~\cite{Moretti:2006sv,Ellwanger:2010nf,
Cao:2011pg,AlbornozVasquez:2012foo,Ellwanger:2012ke,Boudjema:2012in,Schmidt-Hoberg:2012dba,Badziak:2013bda,Badziak:2013gla,Barbieri:2013nka,Fan:2013gjf,
Potter:2015wsa,Ellwanger:2015uaz,Cao:2016uwt,Cao:2019ofo,Cao:2023gkc,Cao:2024axg} indicates that the diphoton rate may be several times larger than its SM prediction for the same scalar mass in the next-to-minimal supersymmetric standard model (NMSSM). In the Two-Higgs doublet model (2HDM) with an additional real singlet (N2HDM), the possibilities of explaining the observed excesses were studied in Refs.~\cite{Biekotter:2019kde,Biekotter:2021ovi,Biekotter:2021qbc,Heinemeyer:2021msz,Biekotter:2022jyr,Biekotter:2022abc,Biekotter:2023jld,
Azevedo:2023zkg,Aguilar-Saavedra:2023vpd}. The authors of Ref.~\cite{Sachdeva:2019hvk} explored the viability of the radion mixed Higgs to be 125 GeV along with the presence of a light radion which can account for the CMS diphoton excess well in the Higgs radion mixing model. Considering the one-loop corrections to the neutral scalar masses of the $\mu\nu$SSM, the authors of Refs.~\cite{Biekotter:2017xmf,Biekotter:2019gtq} demonstrated how the $\mu\nu$SSM can simultaneously accommodate two excesses measured at the LEP and LHC at the $1\sigma$ level. Based on the analysis in Ref.~\cite{Ashanujjaman:2023etj}, extending the scalar sector with a $SU(2)_L$ triplet (the hypercharge $Y=0$) can well provide the origin of the $95\;{\rm GeV}$ excesses. Whether certain model realizations could simultaneously explain the two excesses while being in agreement with all other Higgs boson related limits and measurements was reviewed in Refs.~\cite{Azatov:2012bz,Heinemeyer:2018wzl}.

In this work, we investigate the diphoton and $b\bar b$ excesses in the $B-L$ supersymmetric model (B-LSSM)~\cite{Khalil:2008ps,Elsayed:2011de,Elsayed:2012ec,Abdallah:2016vcn,Khalil:2015naa,DelleRose:2017uas,Yang:2020bmh,Yang:2021duj,Abdelalim:2020xfk}\footnote{Eq.~(\ref{eq2}) shows obviously that the ditau excess has large experimental uncertainties, combined with the analysis in Ref.~\cite{Iguro:2022dok} which indicates the CP-even dominated Higgs state has difficulties in describing the ditau excess, we focus on the diphoton and $b\bar b$ excesses in this work.}. Compared with the minimal supersymmetric standard model (MSSM), the gauge symmetry of the B-LSSM is extended by an additional $U(1)_{B-L}$ gauge group, i.e. the full gauge structure is $SU(3)_C\otimes SU(2)_L \otimes U(1)_Y\otimes U(1)_{B-L}$. In addition to the MSSM superfields, two singlet chiral Higgs and three right-handed neutrinos with nonzero $U(1)_{B-L}$ charge are introduced in the B-LSSM. The tiny neutrino masses can be acquired naturally through the so-called type I seesaw mechanism when the $U(1)_{B-L}$ symmetry is broken spontaneously by the two newly introduced singlet Higgs. The mixings between the SM-like Higgs and the B-LSSM specific Higgs states are generated through the gauge kinetic mixing effect which arises in the NP models with two Abelian groups. It indicates that the rich Higgs sector and Higgs mixing effect provide great potential to explain the $95\;{\rm GeV}$ excesses. We take one of the B-LSSM specific CP-even Higgs states as the lightest Higgs to explore whether the B-LSSM can account for the low-mass diphoton and $b\bar b$ excesses. It is known that the loop corrections from top and stop quarks are crucial to the SM-like Higgs boson mass in the supersymmetric models, the two-loop effective potential corrections from top and stop quarks to the CP-even neutral Higgs boson masses are considered in the calculations.

The paper is organized as follows. The scalar sector of the B-LSSM and the two-loop effective potential corrections to the CP-even neutral Higgs mass matrix are given in Sec.~\ref{sec2}. The numerical results are presented and analyzed in Sec.~\ref{sec3}. Finally, a summary is made in Sec.~\ref{sec4}. The tedious formulae are collected in the appendix.

\section{The B-LSSM and Higgs boson mass with two-loop effective potential corrections\label{sec2}}

The gauge group of B-LSSM is extended by an additional $U(1)_{B-L}$ local gauge group, and a new $Z'$ gauge boson is introduced correspondingly. The chiral scalar superfields in the B-LSSM are
\begin{eqnarray}
&&H_d=\left(\begin{array}{c}H_d^1\\ H_d^2\end{array}\right)\sim(1,2,-1/2,0),\;H_u=\left(\begin{array}{c}H_u^1\\ H_u^2\end{array}\right)\sim(1,2,1/2,0),\nonumber\\
&&\eta\sim(1,1,0,-1),\;\;\bar\eta\sim(1,1,0,1),
\end{eqnarray}
where the charges in the brackets correspond to $SU(3)_C,\; SU(2)_L,\; U(1)_Y,\; U(1)_{B-L}$, respectively. The local gauge symmetry $SU(2)_L \otimes U(1)_Y\otimes U(1)_{B-L}$ breaks down to the electromagnetic symmetry $U(1)_{\rm em}$ as the Higgs fields receive nonzero vacuum expectation values (VEV):
\begin{eqnarray}
&&H_d^1=\frac{1}{\sqrt2}(v_d+\phi_d+i\sigma_d),
\qquad\; H_u^2=\frac{1}{\sqrt2}(v_u+\phi_u+i\sigma_u),\nonumber\\
&&\eta=\frac{1}{\sqrt2}(u_\eta+\phi_\eta+i\sigma_\eta),
\qquad\;\quad\;\bar\eta=\frac{1}{\sqrt2}(u_{\bar\eta}+\phi_{\bar\eta}+i\sigma_{\bar\eta})\;.
\end{eqnarray}
For convenience, we can define $v^2=v_d^2+v_u^2$, $u^2=u_\eta^2+u_{\bar\eta}^2$, $\tan\beta=v_u/v_d$, $\tan\beta'=u_{\bar\eta}/u_\eta$. The tree-level scalar potential in the B-LSSM can be written as~\cite{Yang:2018guw}
\begin{eqnarray}
&&V^{(0)}=(m_{H_d}^2+\mu^2) H_d^2+(m_{H_u}^2+\mu^2) H_u^2+(m_{\eta}^2+\mu_\eta^2) \eta^2+(m_{\bar\eta}^2+\mu_\eta^2) \bar\eta^2\nonumber\\
&&\qquad\;-2B_\mu H_d H_u-2B_\eta\eta\bar\eta+\frac{1}{8}(g_1^2+g_2^2+g_{YB}^2)(H_d^2-H_u^2)^2+\frac{1}{2}g_B^2(\eta^2-\bar\eta^2)^2\nonumber\\
&&\qquad\;+\frac{1}{2}g_Bg_{YB}(H_d^2-H_u^2)(\eta^2-\bar\eta^2),\label{eq6}
\end{eqnarray}
where $g_B$ is the gauge coupling constant corresponding to the $U(1)_{B-L}$ gauge group, $g_{YB}$ is the coupling constant arising from the kinetic mixing effect. For the one-loop and two-loop effective potential, the dominant contributions come from top and stop quarks. In the B-LSSM, the top quark mass reads
\begin{eqnarray}
&&m_t=\frac{1}{\sqrt2}Y_t v_u.
\end{eqnarray}
The stop quark mass matrix can be written as
\begin{eqnarray}
&&m_{\tilde t}^2=
\left(\begin{array}{cc}m_{\tilde {t}_L}^2,&\frac{Y_t}{\sqrt2}(v_u A_t-v_d\mu)\\ \frac{Y_t}{\sqrt2}(v_u A_t-v_d\mu),&m_{\tilde {t}_ R}^2\end{array}\right),
\end{eqnarray}
where
\begin{eqnarray}
&&m_{\tilde {t}_L}^2=\frac{1}{24}\Big[(g_1^2-3g_2^2+g_{YB}^2+g_Bg_{YB})(v_u^2-v_d^2)+2g_B(g_B+g_{YB})(u_{\bar\eta}^2-
u_\eta^2)\Big]\nonumber\\
&&\qquad\quad+m_{\tilde q}^2+\frac{1}{2}v_u^2Y_t^2,\nonumber\\
&&m_{\tilde {t}_R}^2=\frac{1}{24}\Big[(4g_1^2+4g_{YB}^2+g_Bg_{YB})(v_d^2-v_u^2)+2g_B(g_B+4g_{YB})(u_{\eta}^2-
u_{\bar\eta}^2)\Big]\nonumber\\
&&\qquad\quad+m_{\tilde u}^2+\frac{1}{2}v_u^2Y_t^2.
\end{eqnarray}
On the mass eigenstates, the physical stop masses can be written as
\begin{eqnarray}
&&M_{\tilde {t}_{1,2}}^2=\frac{1}{2}(m_{\tilde {t}_L}^2+m_{\tilde {t}_R}^2)\pm\Big[\frac{1}{4}(m_{\tilde {t}_L}^2-m_{\tilde {t}_R}^2)^2+\frac{1}{2}Y_t^2(v_u A_t-v_d\mu)^2\Big]^{1/2}.
\end{eqnarray}

The effective potential is
\begin{eqnarray}
&&V_{\rm eff}=V^{(0)}+\Delta V=V^{(0)}+V^{(1)}+V^{(2)},\label{eq11}
\end{eqnarray}
with~\cite{Zhang:1998bm,Espinosa:1999zm,Degrassi:2001yf}
\begin{eqnarray}
&&V^{(1)}=-\frac{3}{16\pi^2}(m_t^2)^2(\log \frac{m_t^2}{Q^2}-\frac{3}{2})+\frac{3}{32\pi^2}\sum_{i=1}^2(M_{\tilde t_i}^2)^2(\log \frac{M_{\tilde t_i}^2}{Q^2}-\frac{3}{2}),\nonumber\\
&&V^{(2)}=\frac{\alpha_s}{16 \pi^3}\Big\{2J(m_t^2,m_t^2)-4m_t^2 I(m_t^2,m_t^2,0)+\Big[2M_{\tilde t_1}^2 I(M_{\tilde t_1}^2,M_{\tilde t_1}^2,0)+2L(M_{\tilde t_1}^2,M_{\tilde g}^2,m_t^2)\nonumber\\
&&\qquad\quad-4m_tM_{\tilde g} S_{2\bar\theta} I(M_{\tilde t_1}^2,M_{\tilde g}^2,m_t^2)+\frac{1}{2}(1+2C_{2\bar\theta}^2)J(M_{\tilde t_1}^2,M_{\tilde t_1}^2)+\frac{S_{2\bar\theta}^2}{2}J(M_{\tilde t_1}^2,M_{\tilde t_2}^2)\nonumber\\
&&\qquad\quad+(M_{\tilde t_1}\leftrightarrow M_{\tilde t_2},S_{2\bar\theta}\leftrightarrow-S_{2\bar\theta})\Big]\Big\},\label{eq12}
\end{eqnarray}
where $V^{(1)}$, $V^{(2)}$ denote the one-loop and two-loop effective potential, respectively. In Eq.~(\ref{eq12}), $M_{\tilde g}$ is the mass of sgluon and
\begin{eqnarray}
&&S_{2\bar\theta}=\frac{2(M_{\tilde t})_{12}}{M_{\tilde t_1}^2-M_{\tilde t_2}^2},
\end{eqnarray}
where $(M_{\tilde t})_{12}=Y_t(v_u A_t-v_d\mu)$. For the two-loop functions $I,\;J$ in Eq.~(\ref{eq12}) can be found in the appendix D of Ref.~\cite{Degrassi:2009yq}.

Combining Eq.~(\ref{eq6}) with Eq.~(\ref{eq11}), we can obtain the vacuum stability conditions
\begin{eqnarray}
&&m_{H_d}^2+\mu^2=-\frac{\partial \Delta V}{v_d\partial \phi_d}-\frac{1}{4}g_Bg_{YB}(v_\eta^2-v_{\bar\eta}^2)+B_\mu \tan\beta-\frac{1}{8}(g_1^2+g_2^2+g_{YB}^2)(v_d^2-v_u^2),\nonumber\\
&&m_{H_u}^2+\mu^2=-\frac{\partial \Delta V}{v_u\partial \phi_u}+\frac{1}{4}g_Bg_{YB}(v_\eta^2-v_{\bar\eta}^2)+B_\mu/\tan\beta+\frac{1}{8}(g_1^2+g_2^2+g_{YB}^2)(v_d^2-v_u^2),\nonumber\\
&&m_{H_\eta}^2+\mu_\eta^2=-\frac{\partial \Delta V}{v_\eta\partial \phi_\eta}-\frac{1}{4}g_Bg_{YB}(v_d^2-v_u^2)+B_\eta\tan\beta'-\frac{1}{2}g_B^2(v_\eta^2-v_{\bar\eta}^2),\nonumber\\
&&m_{H_{\bar\eta}}^2+\mu_\eta^2=-\frac{\partial \Delta V}{v_{\bar\eta}\partial \phi_{\bar\eta}}+\frac{1}{4}g_Bg_{YB}(v_d^2-v_u^2)+B_\eta/\tan\beta'+\frac{1}{2}g_B^2(v_\eta^2-v_{\bar\eta}^2).
\end{eqnarray}
Then the elements of the CP-even Higgs squared mass matrix can be written as
\begin{eqnarray}
&&m_{\phi_d\phi_d}^2=\frac{1}{4}(g_1^2+g_2^2+g_{YB}^2)v_d^2+B_\mu v_u/v_d+\Big[-\frac{1}{\phi_d}\frac{\partial\Delta V}{\partial\phi_d}+\frac{\partial^2\Delta V}{\partial\phi_d\partial\phi_d}\Big]\Big|_{\rm VEV},\nonumber\\
&&m_{\phi_d\phi_u}^2=m_{\phi_u\phi_d}^2=-B_\mu-\frac{1}{4}(g_1^2+g_2^2+g_{YB}^2)v_d v_u+\Big[\frac{\partial^2\Delta V}{\partial\phi_d\partial\phi_u}\Big]\Big|_{\rm VEV},\nonumber\\
&&m_{\phi_d\phi_\eta}^2=m_{\phi_\eta\phi_d}^2=\frac{1}{2}g_Bg_{YB}v_d v_\eta+\Big[\frac{\partial^2\Delta V}{\partial\phi_d\partial\phi_\eta}\Big]\Big|_{\rm VEV},\nonumber\\
&&m_{\phi_d\phi_{\bar\eta}}^2=m_{\phi_{\bar\eta}\phi_d}^2=-\frac{1}{2}g_Bg_{YB}v_d v_{\bar\eta}+\Big[\frac{\partial^2\Delta V}{\partial\phi_d\partial\phi_{\bar\eta}}\Big]\Big|_{\rm VEV},\nonumber\\
&&m_{\phi_u\phi_u}^2=\frac{1}{4}(g_1^2+g_2^2+g_{YB}^2)v_u^2+B_\mu v_d/v_u+\Big[-\frac{1}{\phi_u}\frac{\partial\Delta V}{\partial\phi_u}+\frac{\partial^2\Delta V}{\partial\phi_u\partial\phi_u}\Big]\Big|_{\rm VEV},\nonumber\\
&&m_{\phi_u\phi_\eta}^2=m_{\phi_\eta\phi_u}^2=-\frac{1}{2}g_Bg_{YB}v_u v_\eta+\Big[\frac{\partial^2\Delta V}{\partial\phi_u\partial\phi_\eta}\Big]\Big|_{\rm VEV},\nonumber\\
&&m_{\phi_u\phi_{\bar\eta}}^2=m_{\phi_{\bar\eta}\phi_u}^2=\frac{1}{2}g_Bg_{YB}v_u v_{\bar\eta}+\Big[\frac{\partial^2\Delta V}{\partial\phi_u\partial\phi_{\bar\eta}}\Big]\Big|_{\rm VEV},\nonumber\\
&&m_{\phi_\eta\phi_\eta}^2=g_B^2v_\eta^2+B_\eta v_{\bar\eta}/v_\eta+\Big[-\frac{1}{\phi_\eta}\frac{\partial\Delta V}{\partial\phi_\eta}+\frac{\partial^2\Delta V}{\partial\phi_\eta\partial\phi_\eta}\Big]\Big|_{\rm VEV},\nonumber\\
&&m_{\phi_\eta\phi_{\bar\eta}}^2=m_{\phi_{\bar\eta}\phi_\eta}^2=-B_\eta-g_B^2v_\eta v_{\bar\eta}+\Big[\frac{\partial^2\Delta V}{\partial\phi_\eta\partial\phi_{\bar\eta}}\Big]\Big|_{\rm VEV},\nonumber\\
&&m_{\phi_{\bar\eta}\phi_{\bar\eta}}^2=g_B^2v_{\bar\eta}^2+B_\eta v_\eta/v_{\bar\eta}+\Big[-\frac{1}{\phi_{\bar\eta}}\frac{\partial\Delta V}{\partial\phi_{\bar\eta}}+\frac{\partial^2\Delta V}{\partial\phi_{\bar\eta}\partial\phi_{\bar\eta}}\Big]\Big|_{\rm VEV},\label{eq24}
\end{eqnarray}
where the detailed calculations of the contributions from $\Delta V$ in Eq.~(\ref{eq24}) are collected in the Appendix \ref{A1}. The tree-level squared Higgs mass matrix in the B-LSSM can be read easily in Eq.~(\ref{eq24}). And to illustrate clearly that the rich Higgs sector in the B-LSSM provides the potential to explain the $95 \;{\rm GeV}$ excesses, we calculate the eigenvalues of the tree-level squared Higgs mass matrix under the approximation $v^2\ll u^2$, the analytical results can be written as
\begin{eqnarray}
&&m_{h_1,{\rm tree}}^2\approx \frac{2 B_{\eta}(\tan^2\beta'-1)^2}{1+(\tan^2\beta'+1)B_{\eta}/(g_B^2u^2)},\quad m_{h_2,{\rm tree}}^2\approx \frac{(g_1^2+g_2^2+g_{YB}^2)(\tan^2\beta-1)^2v^2}{4(\tan^2\beta+1)^2},\nonumber\\
&&m_{h_3,{\rm tree}}^2\approx \frac{B_\mu}{\tan\beta}(\tan^2\beta+1),\quad m_{h_4,{\rm tree}}^2\approx 2(\frac{g_B^2u^2}{\tan^2\beta+1}+B_{\eta}),\label{eqadd1}
\end{eqnarray}
where the terms proportional to $\mathcal{O}[(\tan^2\beta'-1)^3]$ are neglected for simplicity, $m_{h_i,{\rm tree}}<m_{h_j,{\rm tree}}$ for $i<j$, and $m_{h_1,{\rm tree}}$, $m_{h_4,{\rm tree}}$ are the tree-level masses of the states dominated by the two specific scalar singlets in the B-LSSM, $m_{h_2,{\rm tree}}$, $m_{h_3,{\rm tree}}$ are the tree-level masses of the states dominated by the two scalar doublets which correspond to the two scalar doublets in the MSSM. As shown in Eq.~(\ref{eqadd1}), $m_{h_2,{\rm tree}}$ corresponds to the mass of SM-like Higgs, and $m_{h_3,{\rm tree}}\approx m_A\gtrsim 380\;{\rm GeV}$~\cite{PDG} with $m_A$ denoting the CP-odd Higgs mass in the MSSM, which also indicates the $95 \;{\rm GeV}$ Higgs can not be achieved in the MSSM. However, the expression of $m_{h_1,{\rm tree}}$ in Eq.~(\ref{eqadd1}) shows that the $95 \;{\rm GeV}$ Higgs can be obtained in the B-LSSM when $\tan\beta'-1$ approaches to $0$ or $B_{\eta}$ is small. Combined with the tree-level mixing with the SM-like Higgs as presented in Eq.~(\ref{eq24}), the B-LSSM has great potential to explain the observed $95 \;{\rm GeV}$ excesses, while the excesses can not be explained in the frame work of MSSM~\cite{Bechtle:2016kui}.

On the basis $(\phi_d,\phi_u,\phi_\eta,\phi_{\bar \eta})$, the squared mass matrix of CP-even Higgs can be written as
\begin{eqnarray}
&&M_\phi^2=\left(\begin{array}{*{20}{c}}
m_{\phi_d\phi_d}^2 & m_{\phi_d\phi_u}^2 & m_{\phi_d\phi_\eta}^2 & m_{\phi_d\phi_{\bar\eta}}^2\\ [6pt]
m_{\phi_d\phi_u}^2 & m_{\phi_u\phi_u}^2 & m_{\phi_u\phi_\eta}^2 & m_{\phi_u\phi_{\bar\eta}}^2\\ [6pt]
m_{\phi_d\phi_\eta}^2 & m_{\phi_u\phi_\eta}^2 & m_{\phi_\eta\phi_\eta}^2 & m_{\phi_\eta\phi_{\bar\eta}}^2\\ [6pt]
m_{\phi_d\phi_{\bar\eta}}^2 & m_{\phi_u\phi_{\bar\eta}}^2 & m_{\phi_\eta\phi_{\bar\eta}}^2 & m_{\phi_{\bar\eta}\phi_{\bar\eta}}^2
\end{array}\right).
\end{eqnarray}
In order to present the mass matrices of the Higgs fields in a physical way, we rotate the fields $H_d$, $H_u$, $\eta$, $\bar\eta$ as
\begin{eqnarray}
&&H_1=\cos\beta H_u+\sin\beta H_d,\;\;H_2=\cos\beta H_d-\sin\beta H_u,\nonumber\\
&&H_3=\cos\beta' \bar\eta+\sin\beta' \eta,\;\;H_4=\cos\beta' \eta-\sin\beta' \bar\eta,
\end{eqnarray}
where $H_2$ corresponds to the SM Higgs doublet~\cite{Cao:2016uwt}. Then on the basis $(\phi_1,\phi_2,\phi_3,\phi_4)$, the squared mass matrix of CP-even Higgs can be written as
\begin{eqnarray}
&&M_h^2=U\cdot M_\phi^2\cdot U^\dagger,\label{eq18}
\end{eqnarray}
where
\begin{eqnarray}
&&U=\left(\begin{array}{*{20}{c}}
\sin\beta & \cos\beta & 0 & 0\\ [6pt]
\cos\beta & -\sin\beta & 0 & 0\\ [6pt]
0 & 0 & \sin\beta' & \cos\beta'\\ [6pt]
0 & 0 & \cos\beta' & -\sin\beta'
\end{array}\right),\\
&&\phi_1=\cos\beta \phi_u+\sin\beta \phi_d,\;\;\phi_2=\cos\beta \phi_d-\sin\beta \phi_u,\nonumber\\
&&\phi_3=\cos\beta' \phi_{\bar\eta}+\sin\beta' \phi_\eta,\;\;\phi_4=\cos\beta' \phi_\eta-\sin\beta' \phi_{\bar\eta}.
\end{eqnarray}
The squared mass matrix $M_h^2$ can be diagonalized by the unitary matrix $Z_h$ as
\begin{eqnarray}
&&{\rm diag}(m_{h_1}^2,m_{h_2}^2,m_{h_3}^2,m_{h_4}^2)=Z_h\cdot M_h^2\cdot Z_h^\dagger,
\end{eqnarray}
where $h_i,(i=1,2,3,4)$ denote the mass eigenstates.

As defined in the introduction sector, the diphoton and $b\bar b$ signal strength are
\begin{eqnarray}
&&\mu(h_{95})_{\gamma\gamma}=\frac{\sigma(gg\rightarrow h^{\rm NP}_{95}){\rm BR}(h^{\rm NP}_{95}\rightarrow \gamma\gamma)}{\sigma(gg\rightarrow h^{\rm SM}_{95}){\rm BR}(h^{\rm SM}_{95}\rightarrow \gamma\gamma)},\nonumber\\
&&\mu(h_{95})_{bb}=\frac{\sigma(Z^*\rightarrow Z h^{\rm NP}_{95}){\rm BR}(h^{\rm NP}_{95}\rightarrow b\bar b)}{\sigma(Z^*\rightarrow Z h^{\rm SM}_{95}){\rm BR}(h^{\rm SM}_{95}\rightarrow b\bar b)},
\end{eqnarray}
where~\cite{Djouadi:1997yw}
\begin{eqnarray}
&&\Gamma^{\rm SM}_{\rm tot}\approx 0.0026\;{\rm GeV},\nonumber\\
&&{\rm BR}(h^{\rm SM}_{95}\rightarrow \gamma\gamma)\approx 1.4\times10^{-3},\;\;{\rm BR}(h^{\rm SM}_{95}\rightarrow b\bar b)\approx0.81.
\end{eqnarray}
Since top quark makes the dominant contributions to the Higgs production, the cross section for $gg\to h^{NP}_{95}$ and $\sigma(Z^*\rightarrow Z h^{\rm NP}_{95})$ can be estimated by~\cite{Cao:2016uwt}
\begin{eqnarray}
&&\sigma(gg\rightarrow h^{\rm NP}_{95})\approx C_{h_{95}uu}^2\sigma(gg\rightarrow h^{\rm SM}_{95}),\nonumber\\
&&\sigma(Z^*\rightarrow Z h^{\rm NP}_{95})\approx C_{h_{95}VV}^2\sigma(Z^*\rightarrow Z h^{\rm SM}_{95}),
\end{eqnarray}
where the coefficients $C_{h_{95}uu}$, $C_{h_{95}VV}$ are the normalized couplings of $95\;{\rm GeV}$ Higgs in NP models with up-quarks and gauge bosons respectively (in units of the corresponding SM couplings). The branch ratios for $h^{NP}_{95}\to\gamma\gamma$ or $b\bar b$ can be estimated by~\cite{Cao:2016uwt}
\begin{eqnarray}
&&{\rm BR}(h^{\rm NP}_{95}\rightarrow \gamma\gamma)\approx C_{h_{95}uu}^2 {\rm BR}(h^{\rm SM}_{95}\rightarrow \gamma\gamma)\frac{\Gamma^{\rm SM}_{\rm tot}}{\Gamma^{\rm NP}_{\rm tot}},\nonumber\\
&&{\rm BR}(h^{\rm NP}_{95}\rightarrow b\bar b)\approx C_{h_{95}dd}^2 {\rm BR}(h^{\rm SM}_{95}\rightarrow b\bar b)\frac{\Gamma^{\rm SM}_{\rm tot}}{\Gamma^{\rm NP}_{\rm tot}},
\end{eqnarray}
where the coefficients $C_{h_{95}dd}$ is the normalized couplings of
$95\;{\rm GeV}$ Higgs in NP models with down-quarks, and~\cite{Djouadi:1997yw}
\begin{eqnarray}
&&\Gamma^{\rm NP}_{\rm tot}\approx C_{h_{95}dd}^2(\Gamma^{\rm SM}_{b\bar b}+\Gamma^{\rm SM}_{\tau\bar \tau})+C_{h_{95}uu}^2(\Gamma^{\rm SM}_{c\bar c}+\Gamma^{\rm SM}_{gg})+C_{h_{95}VV}^2\Gamma^{\rm SM}_{WW^*},\nonumber\\
&&{\rm BR}(h^{\rm SM}_{95}\rightarrow \tau\tau)\approx 0.082,\;{\rm BR}(h^{\rm SM}_{95}\rightarrow c\bar c)\approx 0.037,{\rm BR}(h^{\rm SM}_{95}\rightarrow gg)\approx 0.058,\nonumber\\
&&{\rm BR}(h^{\rm SM}_{95}\rightarrow WW^*)\approx 0.012.
\end{eqnarray}
In the B-LSSM, we have
\begin{eqnarray}
&&C_{h_{95}uu}=C_{h_1uu}=Z_{h,12}\frac{v}{v_u},\;\;C_{h_{95}dd}=C_{h_1 dd}=Z_{h,11}\frac{v}{v_d},\nonumber\\
&&C_{h_{95}VV}=C_{h_1VV}=Z_{h,11}\frac{v_d}{v}+Z_{h,12}\frac{v_u}{v}.
\end{eqnarray}

Particle Data Group collects various signal strengths of $125\;{\rm GeV}$ Higgs, where the results with high precision read ~\cite{PDG}
\begin{eqnarray}
&&\mu(h_{125})_{\gamma\gamma} = \frac{\sigma(gg \to h_{125}^{\rm NP}){\rm BR}(h_{125}^{\rm NP} \to \gamma\gamma)}{\sigma(gg \to h_{125}^{\rm SM}){\rm BR}(h_{125}^{\rm SM} \to \gamma\gamma)} = 1.10 \pm 0.06, \nonumber\\
&&\mu(h_{125})_{WW^*} = \frac{\sigma(gg \to h_{125}^{\rm NP}){\rm BR}(h_{125}^{\rm NP} \to WW^*)}{\sigma(gg \to h_{125}^{\rm SM}){\rm BR}(h_{125}^{\rm SM} \to WW^*)} = 1.00 \pm 0.08, \nonumber\\
&&\mu(h_{125})_{ZZ^*} = \frac{\sigma(gg \to h_{125}^{\rm NP}){\rm BR}(h_{125}^{\rm NP} \to ZZ^*)}{\sigma(gg \to h_{125}^{\rm SM}){\rm BR}(h_{125}^{\rm SM} \to ZZ^*)} = 1.02 \pm 0.08, \nonumber \\
&&\mu(h_{125})_{b\bar{b}} = \frac{\Gamma(h_{125}^{\rm NP} \to VV^*){\rm BR}(h_{125}^{\rm NP} \to b\bar{b})}{\Gamma(h_{125}^{\rm SM} \to VV^*){\rm BR}(h_{125}^{\rm SM} \to b\bar{b})} = 0.99 \pm 0.12, \nonumber \\
&&\mu(h_{125})_{\tau\bar{\tau}} = \frac{\Gamma(h_{125}^{\rm NP} \to VV^*){\rm BR}(h_{125}^{\rm NP} \to \tau\bar{\tau})}{\Gamma(h_{125}^{\rm SM} \to VV^*){\rm BR}(h_{125}^{\rm SM} \to \tau\bar{\tau})} = 0.91 \pm 0.09, \label{125s}
\end{eqnarray}
and $VV^*$ denotes $WW^*$ or $ZZ^*$. Since top quark makes the dominant contributions to the Higgs production, $\sigma(gg \to h_{125}^{\rm NP})$ and $\Gamma(h_{125}^{\rm NP} \to VV^*)$ can be estimated by ~\cite{Cao:2016uwt}
\begin{eqnarray}
&&\sigma(gg \to h_{125}^{\rm NP}) \approx C_{h_{125}uu}^2 \sigma(gg \to h_{125}^{\rm SM}), \nonumber \\
&&\Gamma(h_{125}^{\rm NP} \to VV^*) \approx C_{h_{125}VV}^2 \Gamma(h_{125}^{\rm SM} \to VV^*),
\end{eqnarray}
where the coefficients $C_{h_{125}uu}$, $C_{h_{125}VV}$ are the normalized couplings of 125 GeV Higgs in NP models with up-quarks and gauge bosons respectively (in units of the corresponding SM couplings). For Higgs boson in the SM, we have
\begin{eqnarray}
&&\Gamma_{\rm tot}^{h_{125}^{\rm SM}} \approx 0.0037~{\rm GeV},~ {\rm BR}(h_{125}^{\rm SM} \to b\bar{b}) \approx 0.53,~ {\rm BR}(h_{125}^{\rm SM} \to WW^*) \approx 0.26, \nonumber \\
&&{\rm BR}(h_{125}^{\rm SM} \to \tau\bar{\tau}) \approx 0.06,~ {\rm BR}(h_{125}^{\rm SM} \to ZZ^*) \approx 0.028,~ {\rm BR}(h_{125}^{\rm SM} \to \gamma\gamma) \approx 0.0025.
\end{eqnarray}

The branch ratios for $h_{125}^{\rm NP} \to \gamma\gamma$, $h_{125}^{\rm NP} \to WW^*$, $h_{125}^{\rm NP} \to ZZ^*$, $h_{125}^{\rm NP} \to b\bar{b}$, $h_{125}^{\rm NP} \to \tau\bar{\tau}$ can be estimated by
\begin{eqnarray}
&&{\rm BR}(h_{125}^{\rm NP} \to \gamma\gamma) \approx C_{h_{125}uu}^2 {\rm BR}(h_{125}^{\rm SM} \to \gamma\gamma) \frac{\Gamma_{\rm tot}^{h_{125}^{\rm SM}}}{\Gamma_{\rm tot}^{h_{125}^{\rm NP}}}, \nonumber \\
&&{\rm BR}(h_{125}^{\rm NP} \to VV^*) \approx C_{h_{125}VV}^2 {\rm BR}(h_{125}^{\rm SM} \to VV^*) \frac{\Gamma_{\rm tot}^{h_{125}^{\rm SM}}}{\Gamma_{\rm tot}^{h_{125}^{\rm NP}}}, \nonumber \\
&&{\rm BR}(h_{125}^{\rm NP} \to b\bar{b}) \approx C_{h_{125}dd}^2 {\rm BR}(h_{125}^{\rm SM} \to b\bar{b}) \frac{\Gamma_{\rm tot}^{h_{125}^{\rm SM}}}{\Gamma_{\rm tot}^{h_{125}^{\rm NP}}}, \nonumber \\
&&{\rm BR}(h_{125}^{\rm NP} \to \tau\bar{\tau}) \approx C_{h_{125}dd}^2 {\rm BR}(h_{125}^{\rm SM} \to \tau\bar{\tau}) \frac{\Gamma_{\rm tot}^{h_{125}^{\rm SM}}}{\Gamma_{\rm tot}^{h_{125}^{\rm NP}}},
\end{eqnarray}
where the coefficients $C_{h_{125}dd}$ is the normalized couplings of $125\;{\rm GeV}$ Higgs in NP models with down-quarks, and
\begin{eqnarray}
\Gamma_{\rm tot}^{h_{125}^{\rm NP}} \approx C_{h_{125}dd}^2 (\Gamma_{b\bar{b}}^{h_{125}^{\rm SM}} + \Gamma_{\tau\bar{\tau}}^{h_{125}^{\rm SM}}) + C_{h_{125}uu}^2 \Gamma_{\gamma\gamma}^{h_{125}^{\rm SM}} + C_{h_{95}VV}^2 (\Gamma_{WW^*}^{h_{125}^{\rm SM}} + \Gamma_{ZZ^*}^{h_{125}^{\rm SM}}).
\end{eqnarray}
In the B-LSSM, we have
\begin{eqnarray}
&&C_{h_{125}uu} = C_{h_2uu} = Z_{h,22} \frac{v}{v_u},\quad C_{h_{125}dd} = C_{h_2dd} = Z_{h,21} \frac{v}{v_d}, \nonumber \\
&&C_{h_{125}VV} = Z_{h,21} \frac{v_d}{v} + Z_{h,22} \frac{v_u}{v}.
\end{eqnarray}

\section{Numerical results\label{sec3}}

Taking the lightest Higgs mass at around $95\;{\rm GeV}$ and the next-to-lightest Higgs mass at the measured SM-like Higgs mass~\cite{PDG}
\begin{eqnarray}
&&m_h=125.20\pm0.11{\rm GeV},\label{hma}
\end{eqnarray}
we present the numerical results of the two light Higgs masses, the diphoton and $b\bar b$ signal strengths in this section.

To carry out the numerical evaluations, we take the relevant SM input parameters as $m_W=80.385\;{\rm GeV},\;m_Z=90.1876\;{\rm GeV},\;\alpha_{em}(m_Z)=1/128.9,\;\alpha_s(m_Z)=0.118$. Due to the introducing of $U(1)_{B-L}$ gauge group, a new $Z'$ gauge boson is introduced in the B-LSSM. Refs.~\cite{newZ,20,21} give an upper bound on the ratio between the $Z'$ mass $M_{Z'}\approx\frac{1}{2}g_B u$ and its gauge coupling at $99\%$ CL as
\begin{eqnarray}
&&M_{Z'}/g_{_B}\geq6\;{\rm TeV}\;.\label{eq27}
\end{eqnarray}
For simplicity, we take $g_B=0.5$ in the following analysis, and Eq.~(\ref{eq27}) indicates $M_{Z'}\geq3\;{\rm TeV}$ in this case. As analyzed in Refs.~\cite{Cacciapaglia:2006pk,Yang:2023lne}, the effects of new $Z'$ in the B-LSSM on the couplings of $Z-$charged leptons which are measured at LEP precisely~\cite{PDG}, can be eliminated by redefining the neutral currents involving charged leptons. It indicates that the redefined $Z$ boson in the B-LSSM can coincide with the one measured at the LEP~\cite{PDG}. The LHC experimental data constrains $\tan\beta^{'}<1.5$~\cite{8}. Based on the analysis about $\bar B\rightarrow X_s\gamma$ and $B_s^0\rightarrow \mu^+\mu^-$ in our previous work~\cite{Yang:2018fvw}, we take $m_{H^\pm}=1.5\;{\rm TeV}$. To coincide with the constraints from the direct searches of squarks, sgluon at the LHC~\cite{ATLAS.PRD,CMS.JHEP} and the observed Higgs signal analyzed in Ref.~\cite{add1}, we take $m_{\tilde g}=m_{\tilde q}=m_{\tilde u}=2.5\;{\rm TeV}$ for simplicity. It is well known that squarks can make contributions at the one-loop level to the oblique parameters $S,\;T,\;U$, which suffer strict constraints from the LEP measurements~\cite{PDG}. Hence, the contributions from squarks in the B-LSSM to $S,\;T,\;U$ parameters are calculated and the details are collected in the appendix~\ref{A2}. We verify numerically that the $S,\;T,\;U$ parameters can satisfy the LEP measurements for the chosen squark masses.

\begin{figure}
\setlength{\unitlength}{1mm}
\centering
\includegraphics[width=2.7in]{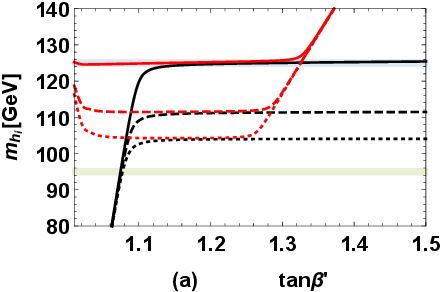}
\vspace{0.5cm}
\includegraphics[width=2.7in]{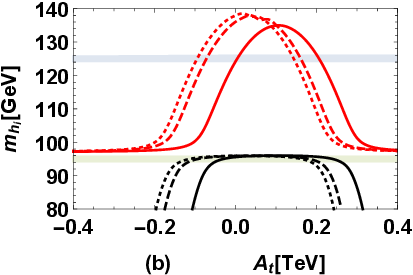}
\vspace{0cm}
\caption[]{Taking $g_{YB}=-0.4$, $M_{Z'}=4.2\;{\rm TeV}$, $B_\eta=(1\;{\rm TeV})^2$, $m_{h_i}$ versus $\tan\beta'$ for $A_t=0.2\;{\rm TeV}$ is plotted in (a), and $m_{h_i}$ versus $A_t$ for $\tan\beta'=1.075$ is plotted in (b). The black, red lines denote the results for $m_{h_1}$, $m_{h_2}$ respectively, the solid, dashed, dotted lines denote the results for $\tan\beta=5,\;15,\;25$ respectively, the green areas denote the range $94\;{\rm GeV}<m_{h_i}<96\;{\rm GeV}$, the gray areas denote the range $124\;{\rm GeV}<m_{h_i}<126\;{\rm GeV}$.}
\label{Fig1}
\end{figure}

Firstly, we investigate the masses of the lightest and next-to-lightest Higgs bosons within the B-LSSM. Taking $g_{YB}=-0.4$, $M_{Z'}=4.2\;{\rm TeV}$, $B_\eta=(1\;{\rm TeV})^2$, we plot $m_{h_i}$ versus $\tan\beta'$ for $A_t=0.2\;{\rm TeV}$ in Fig.~\ref{Fig1} (a), and $m_{h_i}$ versus $A_t$ for $\tan\beta'=1.075$ in Fig.~\ref{Fig1} (b). Where the black, red lines denote the results for $m_{h_1}$, $m_{h_2}$ respectively, the solid, dashed, dotted lines denote the results for $\tan\beta=5,\;15,\;25$ respectively, the green areas denote the range $94\;{\rm GeV}<m_{h_i}<96\;{\rm GeV}$, the gray areas denote the range $124\;{\rm GeV}<m_{h_i}<126\;{\rm GeV}$.
It can be noted from the picture that $\tan\beta'$, $\tan\beta$ and $A_t$ affect the two light Higgs boson masses obviously. Importantly, both the $95\;{\rm GeV}$ Higgs and the SM-like Higgs with a mass of $125\;{\rm GeV}$ are attainable within the B-LSSM. Fig.~\ref{Fig1} (a) demonstrates that $\tan\beta'$ is constrained strictly for $m_{h_1}\approx 95\;{\rm GeV}$ in our chosen parameter space, and the effects of $\tan\beta'$ on $m_{h_i}\;(i=1,2)$ are influenced significantly by the value of $\tan\beta$. This fact can be seen explicitly in Eq.~(\ref{eqadd1}). From Fig.~\ref{Fig1} (b) we can see that $A_t$ is subject to a strict limitation through the requirement of setting $m_{h_2}\approx 125\;{\rm GeV}$.

\begin{figure}
\setlength{\unitlength}{1mm}
\centering
\includegraphics[width=2.7in]{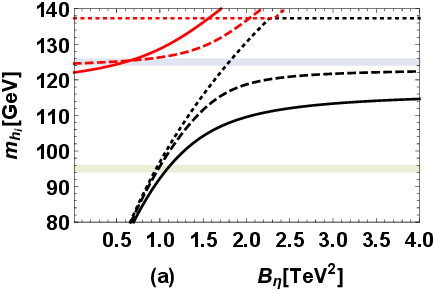}
\vspace{0.5cm}
\includegraphics[width=2.7in]{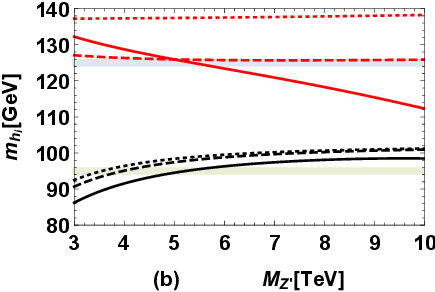}
\vspace{0cm}
\caption[]{Taking $\tan\beta'=1.075$, $\tan\beta=5$, $A_t=0.2\;{\rm TeV}$, $m_{h_i}$ versus $B_\eta$ for $M_{Z'}=4.2\;{\rm TeV}$ is plotted in (a), and $m_{h_i}$ versus $M_{Z'}$ for $B_\eta=(1\;{\rm TeV})^2$ is plotted in (b). The black, red lines denote the results for $m_{h_1}$, $m_{h_2}$ respectively, the solid, dashed, dotted lines denote the results for $g_{YB}=-0.8,\;-0.4,\;0$ respectively, the green areas denote the range $94\;{\rm GeV}<m_{h_i}<96\;{\rm GeV}$, the gray areas denote the range $124\;{\rm GeV}<m_{h_i}<126\;{\rm GeV}$.}
\label{Fig2}
\end{figure}

As shown in Eq.~(\ref{eqadd1}), the lightest Higgs boson is dominated by one of the B-LSSM specific Higgs states, and $B_\eta$, $g_{YB}$, $M_{Z'}$ can affect the theoretical predictions on the two light Higgs boson masses significantly. Then we take $\tan\beta'=1.075$, $\tan\beta=5$, $A_t=0.2\;{\rm TeV}$ and plot $m_{h_i}$ versus $B_\eta$ for $M_{Z'}=4.2\;{\rm TeV}$ in Fig.~\ref{Fig2} (a), and $m_{h_i}$ versus $M_{Z'}$ for $B_\eta=(1\;{\rm TeV})^2$ in Fig.~\ref{Fig2} (b). Where the black, red lines denote the results for $m_{h_1}$, $m_{h_2}$ respectively, the solid, dashed, dotted lines denote the results for $g_{YB}=-0.8,\;-0.4,\;0$ respectively, the green areas denote the range $94\;{\rm GeV}<m_{h_i}<96\;{\rm GeV}$, the gray areas denote the range $124\;{\rm GeV}<m_{h_i}<126\;{\rm GeV}$.

Fig.~\ref{Fig2} (a) shows that both $m_{h_1}$ and $m_{h_2}$ increase as $B_\eta$ increases, and the dotted lines intersect. At $g_{YB}=0$, there is no mixing effects between the SM-like Higgs with the B-LSSM specific Higgs states at the tree level (the fact can be seen obviously in Eq.~(\ref{eq24})). In this scenario, $h_2$ is primarily governed by the SM-like Higgs state and $m_{h_2}$ does not depend on the value of $B_\eta$. Conversely, $m_{h_1}$ is dominated by the B-LSSM specific Higgs state and rises with increasing $B_\eta$. As $B_\eta$ increases, the mass of the B-LSSM specific Higgs state surpasses the mass of the SM-like Higgs state, causing $h_1$ to be dominated by the SM-like Higgs state and $h_2$ to be dominated by the B-LSSM specific Higgs state. This situation results in the intersection of the two dotted lines in Fig.~\ref{Fig2} (a). The three red lines in Fig.~\ref{Fig2} (b) show that $M_{Z'}$ affects $m_{h_2}$ obviously when $g_{YB}=-0.8$, because the mixing effects between the SM-like Higgs with the B-LSSM specific Higgs states are large for large $g_{YB}$ and play important roles on SM-like Higgs boson masses.

Based on the preceding analysis, it is evident that both the $95\;{\rm GeV}$ Higgs boson and the $125\;{\rm GeV}$ SM-like Higgs boson are attainable in the B-LSSM. Due to the masses of the two light Higgs bosons are influenced significantly in a complex manner by $\tan\beta$, $\tan\beta'$, $A_t$, $B_\eta$, $g_{YB}$, $M_{Z'}$, we scan the following parameter space
\begin{eqnarray}
&&\tan\beta=(2,40),\;\tan\beta'=(1,1.5),\;A_t=(-1,1)\;{\rm TeV},\nonumber\\
&&B_\eta=(0.1^2,2^2)\;{\rm TeV}^2,\;g_{YB}=(-1,0),\;M_{Z'}=(3,10)\;{\rm TeV},\label{eq28}
\end{eqnarray}
to comprehensively explore the collective influences of $\tan\beta$, $\tan\beta'$, $A_t$, $B_\eta$, $g_{YB}$, $M_{Z'}$ on the signal strengthes of diphoton and $b\bar b$ events, and keep $94\;{\rm GeV}<m_{h_1}<96\;{\rm GeV}$, $124\;{\rm GeV}<m_{h_2}<126\;{\rm GeV}$ in the scanning. It can be noted in Eq.~(\ref{eq28}) that we take $\tan\beta$ in the range $2<\tan\beta<40$, and the neutral and charged Higgs are limited to be heavier for larger $\tan\beta$ by the LHC searches~\cite{PDG}. For $\tan\beta=40$, the neutral Higgs mass and charged Higgs mass are required to be larger than about $1.3\;{\rm TeV}$, $0.9\;{\rm TeV}$ respectively~\cite{PDG}. In our chosen parameter space, the charged Higgs mass is taken as $m_{H^\pm}=1.5\;{\rm TeV}$ for simplicity as mentioned above, and the two neutral Higgs masses are larger than $1.4\;{\rm TeV}$, which indicate that the chosen parameter space well coincide with the LHC searches.

\begin{figure}
\setlength{\unitlength}{1mm}
\centering
\includegraphics[width=2.7in]{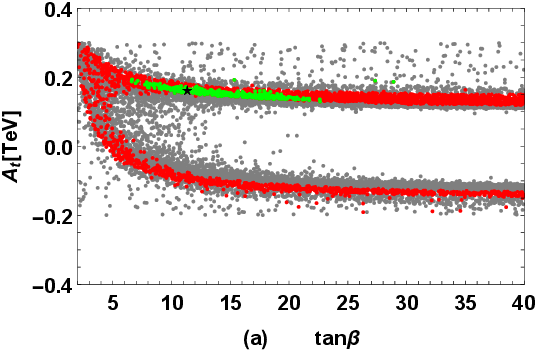}
\vspace{0.5cm}
\includegraphics[width=2.7in]{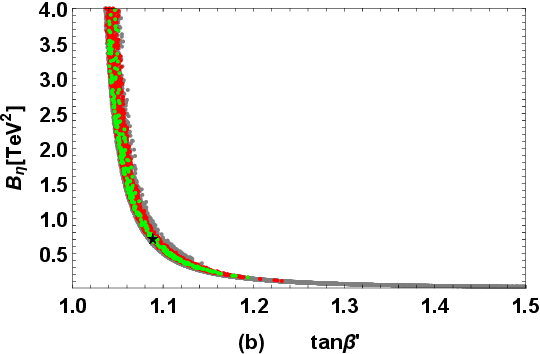}
\vspace{0.5cm}
\includegraphics[width=2.7in]{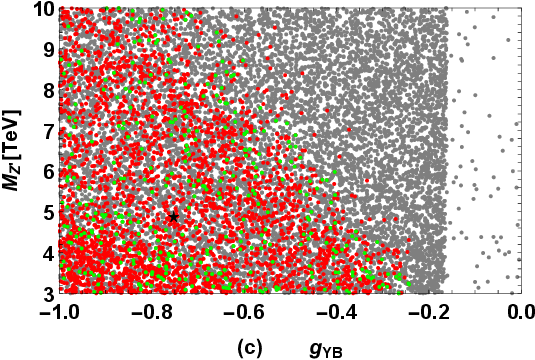}
\vspace{0.5cm}
\includegraphics[width=2.7in]{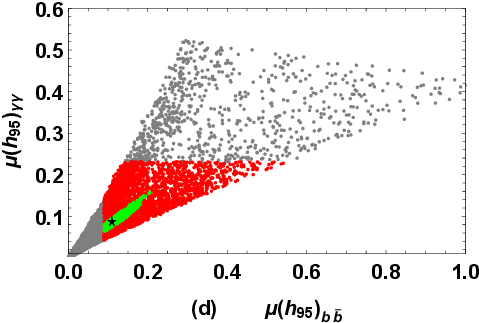}
\vspace{0cm}
\caption[]{Scanning the parameter space in Eq.~(\ref{eq28}) and keeping $94\;{\rm GeV}<m_{h_1}<96\;{\rm GeV}$, $124\;{\rm GeV}<m_{h_2}<126\;{\rm GeV}$ in the scanning, the allowed ranges of $A_t-\tan\beta$ (a), $B_\eta-\tan\beta'$ (b), $M_{Z'}-g_{YB}$ (c) and the obtained $\mu(h_{95})_{\gamma\gamma}-\mu(h_{95})_{b\bar b}$ (d) are plotted. The `black stars' denote the best fit corresponding to $\chi^2=1.5$, the gray points denote the results excluded by considering the $\mu(h_{95})_{\gamma\gamma},\;\mu(h_{95})_{b\bar b}$ in the experimental $2\sigma$ intervals, the red points denote the results excluded by considering the 125 GeV Higgs signal strengths in the experimental $2\sigma$ intervals, and the green points denote the results with $\mu(h_{95})_{\gamma\gamma},\;\mu(h_{95})_{b\bar b}$ and the 125 GeV Higgs signal strengths in the experimental $2\sigma$ intervals.}
\label{Fig3}
\end{figure}

To explore the best fit describing 125 GeV Higgs mass and the diphoton, $b\bar b$ excesses, a $\chi^2$ test is performed. The $\chi^2$ function can be constructed as
\begin{eqnarray}
&&\chi^2=\sum_1 \Big(\frac{O_i^{\rm th}-O_i^{\rm exp}}{\sigma_i^{\rm exp}}\Big)^2,
\end{eqnarray}
where $O_i^{\rm th}$ denotes the $i-$th observable computed theoretically, $O_i^{\rm exp}$ is the corresponding experimental value and $\sigma_i^{\rm exp}$ is the uncertainty in $O_i^{\rm exp}$. Then the allowed ranges of $A_t-\tan\beta$, $B_\eta-\tan\beta'$, $M_{Z'}-g_{YB}$ are plotted in Fig.~\ref{Fig3} (a), Fig.~\ref{Fig3} (b), Fig.~\ref{Fig3} (c) respectively, and the obtained $\mu(h_{95})_{\gamma\gamma}-\mu(h_{95})_{bb}$ is plotted in Fig.~\ref{Fig3} (d). In Fig.~\ref{Fig3}, the gray points denote the results excluded by considering the $\mu(h_{95})_{\gamma\gamma},\;\mu(h_{95})_{b\bar b}$ in the experimental $2\sigma$ intervals, the red points denote the results excluded by considering the 125 GeV Higgs signal strengths in the experimental $2\sigma$ intervals, and the green points denote the results with $\mu(h_{95})_{\gamma\gamma},\;\mu(h_{95})_{b\bar b}$ and the 125 GeV Higgs signal strengths in the experimental $2\sigma$ intervals. The `black stars' denote the best fit corresponding to $\chi^2=1.5$ with the diphoton excess in Eq.~(\ref{eq3}), $b\bar b$ excess in Eq.~(\ref{eq4}) and 125 GeV Higgs mass in Eq.~(\ref{hma}), and 125 GeV Higgs signal strengths in Eq.~(\ref{125s}). The results of the best fit are listed in Tab.~\ref{tab2}. The deviations listed in Tab.~\ref{tab2} indicates that the signal strengths predicted in the B-LSSM are hard to fit the observed results in Eq.~(\ref{eq4}) and Eq.~(\ref{125s}) in the experimental $1\sigma$ intervals, but can fit the observed results in the experimental $2\sigma$ intervals. Hence, the high-precision observations of $\mu(h_{95})_{\gamma\gamma}$ and $\mu(h_{95})_{b\bar b}$ in future may exclude the B-LSSM, or further constraint the parameter space of the model.

\begin{table*}
\begin{tabular*}{\textwidth}{@{\extracolsep{\fill}}llll@{}}
\hline
Observables & $O_i^{\rm th}$ & $O_i^{\rm exp}$ & Deviations in $\%$\\
\hline
$m_{h_1}$ [GeV]                       & 94.83     & 95.00    & 0.12\\
$m_{h_2}$ [GeV]                       & 125.07    & 125.09   & 0.02\\
$\mu(h_{125})_{\gamma\gamma}$         & 1.04      & 1.10     & 5.45\\
$\mu(h_{125})_{WW^*}$                 & 1.04      & 1.00     & 6.00\\
$\mu(h_{125})_{ZZ^*}$                 & 1.04      & 1.02     & 1.96\\
$\mu(h_{125})_{b\bar b}$              & 0.83      & 0.99     & 16.16\\
$\mu(h_{125})_{\tau\bar\tau}$         & 0.83      & 0.91     & 8.79\\
$\mu(h_{95})_{\gamma\gamma}$          & 0.11     & 0.33     & 66.67\\
$\mu(h_{95})_{b\bar b}$               & 0.09     & 0.117    & 23.08\\
\hline
\end{tabular*}
\caption{The results obtained for the best fit corresponding to $\chi^2=\mathbf{1.5}$.}
\label{tab2}
\end{table*}

Fig.~\ref{Fig3} (a) shows that negative $A_t$ is excluded by considering the signal strengths of two light Higgs bosons in the experimental $2\sigma$ intervals, and most of $A_t$ are excluded by the conditions $94\;{\rm GeV}<m_{h_1}<96\;{\rm GeV}$, $124\;{\rm GeV}<m_{h_2}<126\;{\rm GeV}$. The two light Higgs masses prefer large $B_\eta$ for small $\tan\beta'$ and $B_\eta$ is limited strictly for given $\tan\beta'$ as shown in Fig.~\ref{Fig3} (b), because $B_\eta$, $\tan\beta'$ affect the two light Higgs boson masses significantly. In addition, $\tan\beta'$ is limited in the range $1.04\lesssim\tan\beta'\lesssim1.16$ by considering the two light Higgs signal strengths in the experimental $2\sigma$ intervals. Fig.~\ref{Fig3} (c) shows that $g_{YB}$ is limited in the range $g_{YB}\lesssim-0.3$ and $M_{Z'}/|g_{YB}|\lesssim 16.5\;{\rm TeV}$ by considering the two light Higgs signal strengths in the experimental $2\sigma$ intervals, which provides a new bound on the ratio of $Z'$ boson mass with the kinetic mixing parameter $g_{YB}$ if the diphoton and $b\bar b$ excesses are verified in future.

\begin{figure}
\setlength{\unitlength}{1mm}
\centering
\includegraphics[width=2.7in]{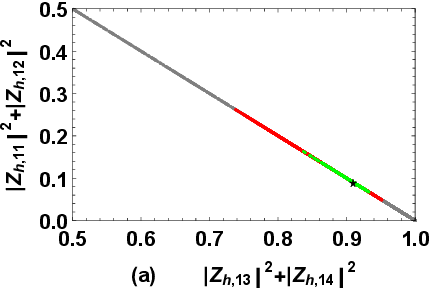}
\vspace{0.5cm}
\includegraphics[width=2.7in]{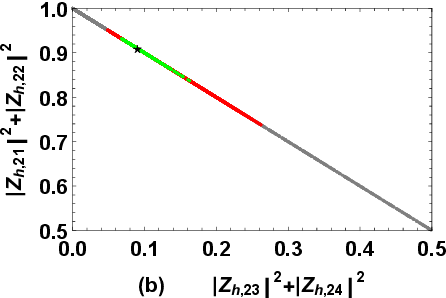}
\vspace{0.5cm}
\caption[]{Scanning the parameter space in Eq.~(\ref{eq28}) and keeping $94\;{\rm GeV}<m_{h_1}<96\;{\rm GeV}$, $124\;{\rm GeV}<m_{h_2}<126\;{\rm GeV}$ in the scanning, the results of $|Z_{h,11}|^2+|Z_{h,12}|^2$ versus $|Z_{h,13}|^2+|Z_{h,14}|^2$ (a), $|Z_{h,21}|^2+|Z_{h,22}|^2$ versus $|Z_{h,23}|^2+|Z_{h,24}|^2$ (b) are plotted. The `black stars' denote the best fit corresponding to $\chi^2=1.5$, the gray points denote the results excluded by considering the $\mu(h_{95})_{\gamma\gamma},\;\mu(h_{95})_{b\bar b}$ in the experimental $2\sigma$ intervals, the red points denote the results excluded by considering the 125 GeV Higgs signal strengths in the experimental $2\sigma$ intervals, and the green points denote the results with $\mu(h_{95})_{\gamma\gamma},\;\mu(h_{95})_{b\bar b}$ and the 125 GeV Higgs signal strengths in the experimental $2\sigma$ intervals.}
\label{Fig4}
\end{figure}

In order to show explicitly the components of the two light Higgs bosons, we plot the results of $|Z_{h,11}|^2+|Z_{h,12}|^2$ versus $|Z_{h,13}|^2+|Z_{h,14}|^2$ in Fig.~\ref{Fig4} (a), and $|Z_{h,21}|^2+|Z_{h,22}|^2$ versus $|Z_{h,23}|^2+|Z_{h,24}|^2$ in Fig.~\ref{Fig4} (b). The definitions of the `black stars', gray points, green points, red points are same to the ones of Fig.~\ref{Fig3}. It can be found from the picture that the lightest $95\;{\rm GeV}$ Higgs is dominated by the new Higgs singlets introduced in the B-LSSM, while the next-to-lightest $125\;{\rm GeV}$ Higgs is dominated by the two Higgs doublets. In addition, it is obvious that the mixing effects between Higgs doublets with B-LSSM specific scalar singlets play important roles in explaining the diphoton and $b\bar b$ excesses at about $95\;{\rm GeV}$, i.e. considering the two loop effective potential corrections is valuable in calculating the two light Higgs boson masses and the scalar mixing effects.

\section{Summary\label{sec4}}

The CMS and LEP experiments have observed the diphoton and $b\bar b$ excesses at a mass about $95\;{\rm GeV}$. Meanwhile, the B-LSSM, which introduces two specific singlet fields, can naturally produce a $95\;{\rm GeV}$ light Higgs boson. Motivated by this, we focus on investigating the masses of the two light Higgs bosons and their corresponding diphoton and $b\bar b$ signal strengths within the B-LSSM framework. The two loop effective potential corrections to the Higgs mass squared matrix are considered in our calculations, because the loop corrections play a crucial roles in determining the SM-like Higgs boson mass within the supersymmetric models. Considering the lightest Higgs boson mass at about $95\;{\rm GeV}$ and the next-to-lightest Higgs boson mass at about $125\;{\rm GeV}$, it is found that the B-LSSM specific Higgs states can account for the observed diphoton and $b\bar b$ excesses around $95\;{\rm GeV}$ simultaneously. As we consider the B-LSSM specific Higgs state as the lightest Higgs boson, the specific parameters $\tan\beta'$, $B_\eta$, $g_{YB}$ and $M_{Z'}$ in the B-LSSM affect the theoretical predictions on the two light Higgs boson masses significantly. The numerical results indicate the B-LSSM is hard to fit these signal strengths in the experimental $1\sigma$ intervals, but the model can reproduce the 125 GeV Higgs signal strengths and the diphoton and $b\bar b$ excesses falling within the experimental $2\sigma$ intervals simultaneously. Furthermore, future high-precision observations of $\mu(h_{95})_{\gamma\gamma}$ and $\mu(h_{95})_{b\bar b}$ could either exclude the B-LSSM or provide further support for its Higgs sector in the B-LSSM. Concurrently, the inclusion of two-loop effective potential corrections is valuable when calculating the masses of the two light Higgs bosons and describing the scalar mixing effects within the Higgs fields, as these effects are crucial for explaining the experimental excesses around $95\;{\rm GeV}$.

\appendix

\section{The two-loop effective potential corrections to the CP-even neutral Higgs mass amtrix.\label{A1}}

Generally, the corrections from the effective potential $\Delta V$ to the CP-even neutral Higgs boson masses can be written as
\begin{eqnarray}
&&\Big[-\frac{1}{\phi_d}\frac{\partial \Delta V}{\partial\phi_d}+\frac{\partial^2 \Delta V}{\partial\phi_d\partial\phi_d}\Big]\Big|_{VEV}=-\frac{1}{v_d}\Big[\frac{\partial \Delta V}{\partial M_{\tilde t_1}^2}\frac{\partial M_{\tilde t_1}^2}{\partial \phi_d}+\frac{\partial \Delta V}{\partial M_{\tilde t_2}^2}\frac{\partial M_{\tilde t_2}^2}{\partial \phi_d}+\frac{\partial \Delta V}{\partial C_{2\bar\theta_t}^2}\frac{\partial C_{2\bar\theta_t}^2}{\partial \phi_d}\Big]\nonumber\\
&&\qquad+\frac{\partial^2 \Delta V}{\partial M_{\tilde t_1}^2\partial M_{\tilde t_1}^2}(\frac{\partial M_{\tilde t_1}^2}{\partial \phi_d})^2+\frac{\partial \Delta V}{\partial M_{\tilde t_1}^2}\frac{\partial^2 M_{\tilde t_1}^2}{\partial \phi_d\partial \phi_d}+2\frac{\partial^2 \Delta V}{\partial M_{\tilde t_1}^2\partial M_{\tilde t_2}^2}\frac{\partial M_{\tilde t_1}^2}{\partial \phi_d}\frac{\partial M_{\tilde t_2}^2}{\partial \phi_d}\nonumber\\
&&\qquad+\frac{\partial^2 \Delta V}{\partial M_{\tilde t_2}^2\partial M_{\tilde t_2}^2}(\frac{\partial M_{\tilde t_2}^2}{\partial \phi_d})^2+\frac{\partial \Delta V}{\partial M_{\tilde t_2}^2}\frac{\partial^2 M_{\tilde t_2}^2}{\partial \phi_d\partial \phi_d}+\frac{\partial^2 \Delta V}{\partial C_{2\bar\theta_t}^2\partial C_{2\bar\theta_t}^2}(\frac{\partial C_{2\bar\theta_t}^2}{\partial \phi_d})^2\nonumber\\
&&\qquad+\frac{\partial \Delta V}{\partial C_{2\bar\theta_t}^2}\frac{\partial^2 C_{2\bar\theta_t}^2}{\partial \phi_d\partial \phi_d}+2\frac{\partial^2 \Delta V}{\partial M_{\tilde t_1}^2\partial C_{2\bar\theta_t}^2}\frac{\partial M_{\tilde t_1}^2}{\partial \phi_d}\frac{\partial C_{2\bar\theta_t}^2}{\partial \phi_d}+2\frac{\partial^2 \Delta V}{\partial C_{2\bar\theta_t}^2\partial M_{\tilde t_2}^2}\frac{\partial C_{2\bar\theta_t}^2}{\partial \phi_d}\frac{\partial M_{\tilde t_2}^2}{\partial \phi_d},\\
&&\Big[-\frac{1}{\phi_u}\frac{\partial \Delta V}{\partial\phi_u}+\frac{\partial^2 \Delta V}{\partial\phi_u\partial\phi_u}\Big]\Big|_{VEV}=-\frac{1}{v_u}\Big[\frac{\partial \Delta V}{\partial m_{t}^2}\frac{\partial m_{t}^2}{\partial \phi_u}+\frac{\partial \Delta V}{\partial M_{\tilde t_1}^2}\frac{\partial M_{\tilde t_1}^2}{\partial \phi_u}+\frac{\partial \Delta V}{\partial M_{\tilde t_2}^2}\frac{\partial M_{\tilde t_2}^2}{\partial \phi_u}\nonumber\\
&&\qquad+\frac{\partial \Delta V}{\partial C_{2\bar\theta_t}^2}\frac{\partial C_{2\bar\theta_t}^2}{\partial \phi_u}\Big]+\frac{\partial^2 \Delta V}{\partial m_{t}^2\partial m_{t}^2}(\frac{\partial m_{t}^2}{\partial \phi_u})^2+\frac{\partial \Delta V}{\partial m_{t}^2}\frac{\partial^2 m_{t}^2}{\partial \phi_u\partial \phi_u}+\frac{\partial^2 \Delta V}{\partial M_{\tilde t_1}^2\partial M_{\tilde t_1}^2}(\frac{\partial M_{\tilde t_1}^2}{\partial \phi_u})^2\nonumber\\
&&\qquad+\frac{\partial \Delta V}{\partial M_{\tilde t_1}^2}\frac{\partial^2 M_{\tilde t_1}^2}{\partial \phi_u\partial \phi_u}+2\frac{\partial^2 \Delta V}{\partial M_{\tilde t_1}^2\partial M_{\tilde t_2}^2}\frac{\partial M_{\tilde t_1}^2}{\partial \phi_u}\frac{\partial M_{\tilde t_2}^2}{\partial \phi_u}+\frac{\partial^2 \Delta V}{\partial M_{\tilde t_2}^2\partial M_{\tilde t_2}^2}(\frac{\partial M_{\tilde t_2}^2}{\partial \phi_u})^2\nonumber\\
&&\qquad+\frac{\partial \Delta V}{\partial M_{\tilde t_2}^2}\frac{\partial^2 M_{\tilde t_2}^2}{\partial \phi_u\partial \phi_u}+2\frac{\partial^2 \Delta V}{\partial m_{t}^2\partial M_{\tilde t_1}^2}\frac{\partial M_{\tilde t_1}^2}{\partial \phi_u}\frac{\partial m_{t}^2}{\partial \phi_u}+2\frac{\partial^2 \Delta V}{\partial m_{t}^2\partial M_{\tilde t_2}^2}\frac{\partial m_{t}^2}{\partial \phi_u}\frac{\partial M_{\tilde t_2}^2}{\partial \phi_u}\nonumber\\
&&\qquad+\frac{\partial^2 \Delta V}{\partial C_{2\bar\theta_t}^2\partial C_{2\bar\theta_t}^2}(\frac{\partial C_{2\bar\theta_t}^2}{\partial \phi_u})^2+\frac{\partial \Delta V}{\partial C_{2\bar\theta_t}^2}\frac{\partial^2 C_{2\bar\theta_t}^2}{\partial \phi_u\partial \phi_u}+2\frac{\partial^2 \Delta V}{\partial m_{t}^2\partial C_{2\bar\theta_t}^2}\frac{\partial m_{t}^2}{\partial \phi_u}\frac{\partial C_{2\bar\theta_t}^2}{\partial \phi_u}\nonumber\\
&&\qquad +2\frac{\partial^2 \Delta V}{\partial M_{\tilde t_1}^2\partial C_{2\bar\theta_t}^2}\frac{\partial M_{\tilde t_1}^2}{\partial \phi_u}\frac{\partial C_{2\bar\theta_t}^2}{\partial \phi_u}+2\frac{\partial^2 \Delta V}{\partial M_{\tilde t_2}^2\partial C_{2\bar\theta_t}^2}\frac{\partial M_{\tilde t_2}^2}{\partial \phi_u}\frac{\partial C_{2\bar\theta_t}^2}{\partial \phi_u},\\
&&\Big[-\frac{1}{\phi_\eta}\frac{\partial \Delta V}{\partial\phi_\eta}+\frac{\partial^2 \Delta V}{\partial\phi_\eta\partial\phi_\eta}\Big]\Big|_{VEV}=-\frac{1}{v_\eta}\Big[\frac{\partial \Delta V}{\partial M_{\tilde t_1}^2}\frac{\partial M_{\tilde t_1}^2}{\partial \phi_\eta}+\frac{\partial \Delta V}{\partial M_{\tilde t_2}^2}\frac{\partial M_{\tilde t_2}^2}{\partial \phi_\eta}+\frac{\partial \Delta V}{\partial C_{2\bar\theta_t}^2}\frac{\partial C_{2\bar\theta_t}^2}{\partial \phi_\eta}\Big]\nonumber\\
&&\qquad+\frac{\partial^2 \Delta V}{\partial M_{\tilde t_1}^2\partial M_{\tilde t_1}^2}(\frac{\partial M_{\tilde t_1}^2}{\partial \phi_\eta})^2+\frac{\partial \Delta V}{\partial M_{\tilde t_1}^2}\frac{\partial^2 M_{\tilde t_1}^2}{\partial \phi_\eta\partial \phi_\eta}+2\frac{\partial^2 \Delta V}{\partial M_{\tilde t_1}^2\partial M_{\tilde t_2}^2}\frac{\partial M_{\tilde t_1}^2}{\partial \phi_\eta}\frac{\partial M_{\tilde t_2}^2}{\partial \phi_\eta}\nonumber\\
&&\qquad+\frac{\partial^2 \Delta V}{\partial M_{\tilde t_2}^2\partial M_{\tilde t_2}^2}(\frac{\partial M_{\tilde t_2}^2}{\partial \phi_\eta})^2+\frac{\partial \Delta V}{\partial M_{\tilde t_2}^2}\frac{\partial^2 M_{\tilde t_2}^2}{\partial \phi_\eta\partial \phi_\eta}+\frac{\partial^2 \Delta V}{\partial C_{2\bar\theta_t}^2\partial C_{2\bar\theta_t}^2}(\frac{\partial C_{2\bar\theta_t}^2}{\partial \phi_\eta})^2\nonumber\\
&&\qquad+\frac{\partial \Delta V}{\partial C_{2\bar\theta_t}^2}\frac{\partial^2 C_{2\bar\theta_t}^2}{\partial \phi_\eta\partial \phi_\eta}+2\frac{\partial^2 \Delta V}{\partial M_{\tilde t_1}^2\partial C_{2\bar\theta_t}^2}\frac{\partial M_{\tilde t_1}^2}{\partial \phi_\eta}\frac{\partial C_{2\bar\theta_t}^2}{\partial \phi_\eta}+2\frac{\partial^2 \Delta V}{\partial C_{2\bar\theta_t}^2\partial M_{\tilde t_2}^2}\frac{\partial C_{2\bar\theta_t}^2}{\partial \phi_\eta}\frac{\partial M_{\tilde t_2}^2}{\partial \phi_\eta},\\
&&\Big[-\frac{1}{\phi_{\bar\eta}}\frac{\partial \Delta V}{\partial\phi_{\bar\eta}}+\frac{\partial^2 \Delta V}{\partial\phi_{\bar\eta}\partial\phi_{\bar\eta}}\Big]\Big|_{VEV}=-\frac{1}{v_{\bar\eta}}\Big[\frac{\partial \Delta V}{\partial M_{\tilde t_1}^2}\frac{\partial M_{\tilde t_1}^2}{\partial \phi_{\bar\eta}}+\frac{\partial \Delta V}{\partial M_{\tilde t_2}^2}\frac{\partial M_{\tilde t_2}^2}{\partial \phi_{\bar\eta}}+\frac{\partial \Delta V}{\partial C_{2\bar\theta_t}^2}\frac{\partial C_{2\bar\theta_t}^2}{\partial \phi_{\bar\eta}}\Big]\nonumber\\
&&\qquad+\frac{\partial^2 \Delta V}{\partial M_{\tilde t_1}^2\partial M_{\tilde t_1}^2}(\frac{\partial M_{\tilde t_1}^2}{\partial \phi_{\bar\eta}})^2+\frac{\partial \Delta V}{\partial M_{\tilde t_1}^2}\frac{\partial^2 M_{\tilde t_1}^2}{\partial \phi_{\bar\eta}\partial \phi_{\bar\eta}}+2\frac{\partial^2 \Delta V}{\partial M_{\tilde t_1}^2\partial M_{\tilde t_2}^2}\frac{\partial M_{\tilde t_1}^2}{\partial \phi_{\bar\eta}}\frac{\partial M_{\tilde t_2}^2}{\partial \phi_{\bar\eta}}\nonumber\\
&&\qquad+\frac{\partial^2 \Delta V}{\partial M_{\tilde t_2}^2\partial M_{\tilde t_2}^2}(\frac{\partial M_{\tilde t_2}^2}{\partial \phi_{\bar\eta}})^2+\frac{\partial \Delta V}{\partial M_{\tilde t_2}^2}\frac{\partial^2 M_{\tilde t_2}^2}{\partial \phi_{\bar\eta}\partial \phi_{\bar\eta}}+\frac{\partial^2 \Delta V}{\partial C_{2\bar\theta_t}^2\partial C_{2\bar\theta_t}^2}(\frac{\partial C_{2\bar\theta_t}^2}{\partial \phi_{\bar\eta}})^2\nonumber\\
&&\qquad+\frac{\partial \Delta V}{\partial C_{2\bar\theta_t}^2}\frac{\partial^2 C_{2\bar\theta_t}^2}{\partial \phi_{\bar\eta}\partial \phi_{\bar\eta}}+2\frac{\partial^2 \Delta V}{\partial M_{\tilde t_1}^2\partial C_{2\bar\theta_t}^2}\frac{\partial M_{\tilde t_1}^2}{\partial \phi_{\bar\eta}}\frac{\partial C_{2\bar\theta_t}^2}{\partial \phi_{\bar\eta}}+2\frac{\partial^2 \Delta V}{\partial C_{2\bar\theta_t}^2\partial M_{\tilde t_2}^2}\frac{\partial C_{2\bar\theta_t}^2}{\partial \phi_{\bar\eta}}\frac{\partial M_{\tilde t_2}^2}{\partial \phi_{\bar\eta}},\\
&&\Big[\frac{\partial^2\Delta V}{\partial\phi_d\partial\phi_u}\Big]\Big|_{VEV}=\frac{\partial^2 \Delta V}{\partial M_{\tilde t_1}^2\partial M_{\tilde t_1}^2}\frac{\partial M_{\tilde t_1}^2}{\partial \phi_d}\frac{\partial M_{\tilde t_1}^2}{\partial \phi_u}+\frac{\partial \Delta V}{\partial M_{\tilde t_1}^2}\frac{\partial^2 M_{\tilde t_1}^2}{\partial \phi_d\partial \phi_u}+\frac{\partial^2 \Delta V}{\partial M_{\tilde t_1}^2\partial m_{t}^2}\frac{\partial M_{\tilde t_1}^2}{\partial \phi_d}\frac{\partial m_{t}^2}{\partial \phi_u}\nonumber\\
&&\qquad+\frac{\partial^2 \Delta V}{\partial M_{\tilde t_1}^2\partial M_{\tilde t_2}^2}\frac{\partial M_{\tilde t_1}^2}{\partial \phi_d}\frac{\partial M_{\tilde t_2}^2}{\partial \phi_u}+\frac{\partial^2 \Delta V}{\partial M_{\tilde t_1}^2\partial C_{2\bar\theta_t}^2}\frac{\partial M_{\tilde t_1}^2}{\partial \phi_d}\frac{\partial C_{2\bar\theta_t}^2}{\partial \phi_u}\nonumber\\
&&\qquad+\frac{\partial^2 \Delta V}{\partial M_{\tilde t_2}^2\partial M_{\tilde t_2}^2}\frac{\partial M_{\tilde t_2}^2}{\partial \phi_d}\frac{\partial M_{\tilde t_2}^2}{\partial \phi_u}+\frac{\partial \Delta V}{\partial M_{\tilde t_2}^2}\frac{\partial^2 M_{\tilde t_2}^2}{\partial \phi_d\partial \phi_u}+\frac{\partial^2 \Delta V}{\partial M_{\tilde t_2}^2\partial m_{t}^2}\frac{\partial M_{\tilde t_2}^2}{\partial \phi_d}\frac{\partial m_{t}^2}{\partial \phi_u}\nonumber\\
&&\qquad+\frac{\partial^2 \Delta V}{\partial M_{\tilde t_2}^2\partial M_{\tilde t_1}^2}\frac{\partial M_{\tilde t_2}^2}{\partial \phi_d}\frac{\partial M_{\tilde t_1}^2}{\partial \phi_u}+\frac{\partial^2 \Delta V}{\partial M_{\tilde t_2}^2\partial C_{2\bar\theta_t}^2}\frac{\partial M_{\tilde t_2}^2}{\partial \phi_d}\frac{\partial C_{2\bar\theta_t}^2}{\partial \phi_u}\nonumber\\
&&\qquad+\frac{\partial^2 \Delta V}{\partial C_{2\bar\theta_t}^2\partial C_{2\bar\theta_t}^2}\frac{\partial C_{2\bar\theta_t}^2}{\partial \phi_d}\frac{\partial C_{2\bar\theta_t}^2}{\partial \phi_u}+\frac{\partial \Delta V}{\partial C_{2\bar\theta_t}^2}\frac{\partial^2 C_{2\bar\theta_t}^2}{\partial \phi_d\partial \phi_u}+\frac{\partial^2 \Delta V}{\partial C_{2\bar\theta_t}^2\partial m_{t}^2}\frac{\partial C_{2\bar\theta_t}^2}{\partial \phi_d}\frac{\partial m_{t}^2}{\partial \phi_u}\nonumber\\
&&\qquad+\frac{\partial^2 \Delta V}{\partial C_{2\bar\theta_t}^2\partial M_{\tilde t_1}^2}\frac{\partial C_{2\bar\theta_t}^2}{\partial \phi_d}\frac{\partial M_{\tilde t_1}^2}{\partial \phi_u}+\frac{\partial^2 \Delta V}{\partial C_{2\bar\theta_t}^2\partial M_{\tilde t_2}^2}\frac{\partial C_{2\bar\theta_t}^2}{\partial \phi_d}\frac{\partial M_{\tilde t_2}^2}{\partial \phi_u},\\
&&\Big[\frac{\partial^2\Delta V}{\partial\phi_d\partial\phi_\eta}\Big]\Big|_{VEV}=\frac{\partial^2 \Delta V}{\partial M_{\tilde t_1}^2\partial M_{\tilde t_1}^2}\frac{\partial M_{\tilde t_1}^2}{\partial \phi_d}\frac{\partial M_{\tilde t_1}^2}{\partial \phi_\eta}+\frac{\partial \Delta V}{\partial M_{\tilde t_1}^2}\frac{\partial^2 M_{\tilde t_1}^2}{\partial \phi_d\partial \phi_\eta}+\frac{\partial^2 \Delta V}{\partial M_{\tilde t_1}^2\partial M_{\tilde t_2}^2}\frac{\partial M_{\tilde t_1}^2}{\partial \phi_d}\frac{\partial M_{\tilde t_2}^2}{\partial \phi_\eta}\nonumber\\
&&\qquad+\frac{\partial^2 \Delta V}{\partial M_{\tilde t_1}^2\partial C_{2\bar\theta_t}^2}\frac{\partial M_{\tilde t_1}^2}{\partial \phi_d}\frac{\partial C_{2\bar\theta_t}^2}{\partial \phi_\eta}+\frac{\partial^2 \Delta V}{\partial M_{\tilde t_2}^2\partial M_{\tilde t_2}^2}\frac{\partial M_{\tilde t_2}^2}{\partial \phi_d}\frac{\partial M_{\tilde t_2}^2}{\partial \phi_\eta}+\frac{\partial \Delta V}{\partial M_{\tilde t_2}^2}\frac{\partial^2 M_{\tilde t_2}^2}{\partial \phi_d\partial \phi_\eta}\nonumber\\
&&\qquad+\frac{\partial^2 \Delta V}{\partial M_{\tilde t_2}^2\partial M_{\tilde t_1}^2}\frac{\partial M_{\tilde t_2}^2}{\partial \phi_d}\frac{\partial M_{\tilde t_1}^2}{\partial \phi_\eta}+\frac{\partial^2 \Delta V}{\partial M_{\tilde t_2}^2\partial C_{2\bar\theta_t}^2}\frac{\partial M_{\tilde t_2}^2}{\partial \phi_d}\frac{\partial C_{2\bar\theta_t}^2}{\partial \phi_\eta}+\frac{\partial^2 \Delta V}{\partial C_{2\bar\theta_t}^2\partial C_{2\bar\theta_t}^2}\frac{\partial C_{2\bar\theta_t}^2}{\partial \phi_d}\frac{\partial C_{2\bar\theta_t}^2}{\partial \phi_\eta}\nonumber\\
&&\qquad+\frac{\partial \Delta V}{\partial C_{2\bar\theta_t}^2}\frac{\partial^2 C_{2\bar\theta_t}^2}{\partial \phi_d\partial \phi_\eta}+\frac{\partial^2 \Delta V}{\partial C_{2\bar\theta_t}^2\partial M_{\tilde t_1}^2}\frac{\partial C_{2\bar\theta_t}^2}{\partial \phi_d}\frac{\partial M_{\tilde t_1}^2}{\partial \phi_\eta}+\frac{\partial^2 \Delta V}{\partial C_{2\bar\theta_t}^2\partial M_{\tilde t_2}^2}\frac{\partial C_{2\bar\theta_t}^2}{\partial \phi_d}\frac{\partial M_{\tilde t_2}^2}{\partial \phi_\eta},\\
&&\Big[\frac{\partial^2\Delta V}{\partial\phi_d\partial\phi_{\bar\eta}}\Big]\Big|_{VEV}=\frac{\partial^2 \Delta V}{\partial M_{\tilde t_1}^2\partial M_{\tilde t_1}^2}\frac{\partial M_{\tilde t_1}^2}{\partial \phi_d}\frac{\partial M_{\tilde t_1}^2}{\partial \phi_{\bar\eta}}+\frac{\partial \Delta V}{\partial M_{\tilde t_1}^2}\frac{\partial^2 M_{\tilde t_1}^2}{\partial \phi_d\partial \phi_{\bar\eta}}+\frac{\partial^2 \Delta V}{\partial M_{\tilde t_1}^2\partial M_{\tilde t_2}^2}\frac{\partial M_{\tilde t_1}^2}{\partial \phi_d}\frac{\partial M_{\tilde t_2}^2}{\partial \phi_{\bar\eta}}\nonumber\\
&&\qquad+\frac{\partial^2 \Delta V}{\partial M_{\tilde t_1}^2\partial C_{2\bar\theta_t}^2}\frac{\partial M_{\tilde t_1}^2}{\partial \phi_d}\frac{\partial C_{2\bar\theta_t}^2}{\partial \phi_{\bar\eta}}+\frac{\partial^2 \Delta V}{\partial M_{\tilde t_2}^2\partial M_{\tilde t_2}^2}\frac{\partial M_{\tilde t_2}^2}{\partial \phi_d}\frac{\partial M_{\tilde t_2}^2}{\partial \phi_{\bar\eta}}+\frac{\partial \Delta V}{\partial M_{\tilde t_2}^2}\frac{\partial^2 M_{\tilde t_2}^2}{\partial \phi_d\partial \phi_{\bar\eta}}\nonumber\\
&&\qquad+\frac{\partial^2 \Delta V}{\partial M_{\tilde t_2}^2\partial M_{\tilde t_1}^2}\frac{\partial M_{\tilde t_2}^2}{\partial \phi_d}\frac{\partial M_{\tilde t_1}^2}{\partial \phi_{\bar\eta}}+\frac{\partial^2 \Delta V}{\partial M_{\tilde t_2}^2\partial C_{2\bar\theta_t}^2}\frac{\partial M_{\tilde t_2}^2}{\partial \phi_d}\frac{\partial C_{2\bar\theta_t}^2}{\partial \phi_{\bar\eta}}+\frac{\partial^2 \Delta V}{\partial C_{2\bar\theta_t}^2\partial C_{2\bar\theta_t}^2}\frac{\partial C_{2\bar\theta_t}^2}{\partial \phi_d}\frac{\partial C_{2\bar\theta_t}^2}{\partial \phi_{\bar\eta}}\nonumber\\
&&\qquad+\frac{\partial \Delta V}{\partial C_{2\bar\theta_t}^2}\frac{\partial^2 C_{2\bar\theta_t}^2}{\partial \phi_d\partial \phi_{\bar\eta}}+\frac{\partial^2 \Delta V}{\partial C_{2\bar\theta_t}^2\partial M_{\tilde t_1}^2}\frac{\partial C_{2\bar\theta_t}^2}{\partial \phi_d}\frac{\partial M_{\tilde t_1}^2}{\partial \phi_{\bar\eta}}+\frac{\partial^2 \Delta V}{\partial C_{2\bar\theta_t}^2\partial M_{\tilde t_2}^2}\frac{\partial C_{2\bar\theta_t}^2}{\partial \phi_d}\frac{\partial M_{\tilde t_2}^2}{\partial \phi_{\bar\eta}},\\
&&\Big[\frac{\partial^2\Delta V}{\partial\phi_u\partial\phi_\eta}\Big]\Big|_{VEV}=\frac{\partial^2 \Delta V}{\partial m_{t}^2\partial M_{\tilde t_1}^2}\frac{\partial m_{t}^2}{\partial \phi_u}\frac{\partial M_{\tilde t_1}^2}{\partial \phi_\eta}+\frac{\partial^2 \Delta V}{\partial m_{t}^2\partial M_{\tilde t_2}^2}\frac{\partial m_{t}^2}{\partial \phi_u}\frac{\partial M_{\tilde t_2}^2}{\partial \phi_\eta}+\frac{\partial^2 \Delta V}{\partial m_{t}^2\partial C_{2\bar\theta_t}^2}\frac{\partial m_{t}^2}{\partial \phi_u}\frac{\partial C_{2\bar\theta_t}^2}{\partial \phi_\eta}\nonumber\\
&&\qquad+\frac{\partial^2 \Delta V}{\partial M_{\tilde t_1}^2\partial M_{\tilde t_1}^2}\frac{\partial M_{\tilde t_1}^2}{\partial \phi_u}\frac{\partial M_{\tilde t_1}^2}{\partial \phi_{\eta}}+\frac{\partial \Delta V}{\partial M_{\tilde t_1}^2}\frac{\partial^2 M_{\tilde t_1}^2}{\partial \phi_u\partial \phi_{\eta}}+\frac{\partial^2 \Delta V}{\partial M_{\tilde t_1}^2\partial M_{\tilde t_2}^2}\frac{\partial M_{\tilde t_1}^2}{\partial \phi_u}\frac{\partial M_{\tilde t_2}^2}{\partial \phi_{\eta}}\nonumber\\
&&\qquad+\frac{\partial^2 \Delta V}{\partial M_{\tilde t_1}^2\partial C_{2\bar\theta_t}^2}\frac{\partial M_{\tilde t_1}^2}{\partial \phi_u}\frac{\partial C_{2\bar\theta_t}^2}{\partial \phi_{\eta}}+\frac{\partial^2 \Delta V}{\partial M_{\tilde t_2}^2\partial M_{\tilde t_2}^2}\frac{\partial M_{\tilde t_2}^2}{\partial \phi_u}\frac{\partial M_{\tilde t_2}^2}{\partial \phi_{\eta}}+\frac{\partial \Delta V}{\partial M_{\tilde t_2}^2}\frac{\partial^2 M_{\tilde t_2}^2}{\partial \phi_u\partial \phi_{\eta}}\nonumber\\
&&\qquad+\frac{\partial^2 \Delta V}{\partial M_{\tilde t_2}^2\partial M_{\tilde t_1}^2}\frac{\partial M_{\tilde t_2}^2}{\partial \phi_u}\frac{\partial M_{\tilde t_1}^2}{\partial \phi_{\eta}}+\frac{\partial^2 \Delta V}{\partial M_{\tilde t_2}^2\partial C_{2\bar\theta_t}^2}\frac{\partial M_{\tilde t_2}^2}{\partial \phi_u}\frac{\partial C_{2\bar\theta_t}^2}{\partial \phi_{\eta}}+\frac{\partial^2 \Delta V}{\partial C_{2\bar\theta_t}^2\partial C_{2\bar\theta_t}^2}\frac{\partial C_{2\bar\theta_t}^2}{\partial \phi_u}\frac{\partial C_{2\bar\theta_t}^2}{\partial \phi_{\eta}}\nonumber\\
&&\qquad+\frac{\partial \Delta V}{\partial C_{2\bar\theta_t}^2}\frac{\partial^2 C_{2\bar\theta_t}^2}{\partial \phi_u\partial \phi_{\eta}}+\frac{\partial^2 \Delta V}{\partial C_{2\bar\theta_t}^2\partial M_{\tilde t_1}^2}\frac{\partial C_{2\bar\theta_t}^2}{\partial \phi_u}\frac{\partial M_{\tilde t_1}^2}{\partial \phi_{\eta}}+\frac{\partial^2 \Delta V}{\partial C_{2\bar\theta_t}^2\partial M_{\tilde t_2}^2}\frac{\partial C_{2\bar\theta_t}^2}{\partial \phi_u}\frac{\partial M_{\tilde t_2}^2}{\partial \phi_{\eta}},\\
&&\Big[\frac{\partial^2\Delta V}{\partial\phi_u\partial\phi_{\bar\eta}}\Big]\Big|_{VEV}=\frac{\partial^2 \Delta V}{\partial m_{t}^2\partial M_{\tilde t_1}^2}\frac{\partial m_{t}^2}{\partial \phi_u}\frac{\partial M_{\tilde t_1}^2}{\partial \phi_\eta}+\frac{\partial^2 \Delta V}{\partial m_{t}^2\partial M_{\tilde t_2}^2}\frac{\partial m_{t}^2}{\partial \phi_u}\frac{\partial M_{\tilde t_2}^2}{\partial \phi_{\bar\eta}}+\frac{\partial^2 \Delta V}{\partial m_{t}^2\partial C_{2\bar\theta_t}^2}\frac{\partial m_{t}^2}{\partial \phi_u}\frac{\partial C_{2\bar\theta_t}^2}{\partial \phi_{\bar\eta}}\nonumber\\
&&\qquad+\frac{\partial^2 \Delta V}{\partial M_{\tilde t_1}^2\partial M_{\tilde t_1}^2}\frac{\partial M_{\tilde t_1}^2}{\partial \phi_u}\frac{\partial M_{\tilde t_1}^2}{\partial \phi_{\bar\eta}}+\frac{\partial \Delta V}{\partial M_{\tilde t_1}^2}\frac{\partial^2 M_{\tilde t_1}^2}{\partial \phi_u\partial \phi_{\bar\eta}}+\frac{\partial^2 \Delta V}{\partial M_{\tilde t_1}^2\partial M_{\tilde t_2}^2}\frac{\partial M_{\tilde t_1}^2}{\partial \phi_u}\frac{\partial M_{\tilde t_2}^2}{\partial \phi_{\bar\eta}}\nonumber\\
&&\qquad+\frac{\partial^2 \Delta V}{\partial M_{\tilde t_1}^2\partial C_{2\bar\theta_t}^2}\frac{\partial M_{\tilde t_1}^2}{\partial \phi_u}\frac{\partial C_{2\bar\theta_t}^2}{\partial \phi_{\bar\eta}}+\frac{\partial^2 \Delta V}{\partial M_{\tilde t_2}^2\partial M_{\tilde t_2}^2}\frac{\partial M_{\tilde t_2}^2}{\partial \phi_u}\frac{\partial M_{\tilde t_2}^2}{\partial \phi_{\bar\eta}}+\frac{\partial \Delta V}{\partial M_{\tilde t_2}^2}\frac{\partial^2 M_{\tilde t_2}^2}{\partial \phi_u\partial \phi_{\bar\eta}}\nonumber\\
&&\qquad+\frac{\partial^2 \Delta V}{\partial M_{\tilde t_2}^2\partial M_{\tilde t_1}^2}\frac{\partial M_{\tilde t_2}^2}{\partial \phi_u}\frac{\partial M_{\tilde t_1}^2}{\partial \phi_{\bar\eta}}+\frac{\partial^2 \Delta V}{\partial M_{\tilde t_2}^2\partial C_{2\bar\theta_t}^2}\frac{\partial M_{\tilde t_2}^2}{\partial \phi_u}\frac{\partial C_{2\bar\theta_t}^2}{\partial \phi_{\bar\eta}}+\frac{\partial^2 \Delta V}{\partial C_{2\bar\theta_t}^2\partial C_{2\bar\theta_t}^2}\frac{\partial C_{2\bar\theta_t}^2}{\partial \phi_u}\frac{\partial C_{2\bar\theta_t}^2}{\partial \phi_{\bar\eta}}\nonumber\\
&&\qquad+\frac{\partial \Delta V}{\partial C_{2\bar\theta_t}^2}\frac{\partial^2 C_{2\bar\theta_t}^2}{\partial \phi_u\partial \phi_{\bar\eta}}+\frac{\partial^2 \Delta V}{\partial C_{2\bar\theta_t}^2\partial M_{\tilde t_1}^2}\frac{\partial C_{2\bar\theta_t}^2}{\partial \phi_u}\frac{\partial M_{\tilde t_1}^2}{\partial \phi_{\bar\eta}}+\frac{\partial^2 \Delta V}{\partial C_{2\bar\theta_t}^2\partial M_{\tilde t_2}^2}\frac{\partial C_{2\bar\theta_t}^2}{\partial \phi_u}\frac{\partial M_{\tilde t_2}^2}{\partial \phi_{\bar\eta}},\\
&&\Big[\frac{\partial^2\Delta V}{\partial\phi_\eta\partial\phi_{\bar\eta}}\Big]\Big|_{VEV}=\frac{\partial^2 \Delta V}{\partial M_{\tilde t_1}^2\partial M_{\tilde t_1}^2}\frac{\partial M_{\tilde t_1}^2}{\partial \phi_\eta}\frac{\partial M_{\tilde t_1}^2}{\partial \phi_{\bar\eta}}+\frac{\partial \Delta V}{\partial M_{\tilde t_1}^2}\frac{\partial^2 M_{\tilde t_1}^2}{\partial \phi_\eta\partial \phi_{\bar\eta}}+\frac{\partial^2 \Delta V}{\partial M_{\tilde t_1}^2\partial M_{\tilde t_2}^2}\frac{\partial M_{\tilde t_1}^2}{\partial \phi_\eta}\frac{\partial M_{\tilde t_2}^2}{\partial \phi_{\bar\eta}}\nonumber\\
&&\qquad+\frac{\partial^2 \Delta V}{\partial M_{\tilde t_1}^2\partial C_{2\bar\theta_t}^2}\frac{\partial M_{\tilde t_1}^2}{\partial \phi_\eta}\frac{\partial C_{2\bar\theta_t}^2}{\partial \phi_{\bar\eta}}+\frac{\partial^2 \Delta V}{\partial M_{\tilde t_2}^2\partial M_{\tilde t_2}^2}\frac{\partial M_{\tilde t_2}^2}{\partial \phi_\eta}\frac{\partial M_{\tilde t_2}^2}{\partial \phi_{\bar\eta}}+\frac{\partial \Delta V}{\partial M_{\tilde t_2}^2}\frac{\partial^2 M_{\tilde t_2}^2}{\partial \phi_\eta\partial \phi_{\bar\eta}}\nonumber\\
&&\qquad+\frac{\partial^2 \Delta V}{\partial M_{\tilde t_2}^2\partial M_{\tilde t_1}^2}\frac{\partial M_{\tilde t_2}^2}{\partial \phi_\eta}\frac{\partial M_{\tilde t_1}^2}{\partial \phi_{\bar\eta}}+\frac{\partial^2 \Delta V}{\partial M_{\tilde t_2}^2\partial C_{2\bar\theta_t}^2}\frac{\partial M_{\tilde t_2}^2}{\partial \phi_\eta}\frac{\partial C_{2\bar\theta_t}^2}{\partial \phi_{\bar\eta}}+\frac{\partial^2 \Delta V}{\partial C_{2\bar\theta_t}^2\partial C_{2\bar\theta_t}^2}\frac{\partial C_{2\bar\theta_t}^2}{\partial \phi_\eta}\frac{\partial C_{2\bar\theta_t}^2}{\partial \phi_{\bar\eta}}\nonumber\\
&&\qquad+\frac{\partial \Delta V}{\partial C_{2\bar\theta_t}^2}\frac{\partial^2 C_{2\bar\theta_t}^2}{\partial \phi_\eta\partial \phi_{\bar\eta}}+\frac{\partial^2 \Delta V}{\partial C_{2\bar\theta_t}^2\partial M_{\tilde t_1}^2}\frac{\partial C_{2\bar\theta_t}^2}{\partial \phi_\eta}\frac{\partial M_{\tilde t_1}^2}{\partial \phi_{\bar\eta}}+\frac{\partial^2 \Delta V}{\partial C_{2\bar\theta_t}^2\partial M_{\tilde t_2}^2}\frac{\partial C_{2\bar\theta_t}^2}{\partial \phi_\eta}\frac{\partial M_{\tilde t_2}^2}{\partial \phi_{\bar\eta}},
\end{eqnarray}
where
\begin{eqnarray}
&&C_{2\bar\theta_t}^2=1-\frac{2|m_{\tilde t_{12}}^2|^2}{(M_{\tilde t_1}^2-M_{\tilde t_2}^2)^2}.
\end{eqnarray}
From the concrete expression of $\Delta V$ defined in Eqs.~(\ref{eq11},~\ref{eq12}), we can obtain
\begin{eqnarray}
&&\frac{\partial V^{(1)}}{\partial m_t^2}=-\frac{3m_t^2}{8\pi^2}(\log\frac{m_t^2}{Q^2}-1),\;\;\frac{\partial^2 V^{(1)}}{\partial m_t^2\partial m_t^2}=-\frac{3}{8\pi^2}\log\frac{m_t^2}{Q^2},\\
&&\frac{\partial V^{(1)}}{\partial M_{\tilde t_1}^2}=\frac{3M_{\tilde t_1}^2}{16\pi^2}(\log\frac{M_{\tilde t_1}^2}{Q^2}-1),\;\;\frac{\partial^2 V^{(1)}}{\partial M_{\tilde t_1}^2\partial M_{\tilde t_1}^2}=\frac{3}{16\pi^2}\log\frac{M_{\tilde t_1}^2}{Q^2},\\
&&\frac{\partial V^{(1)}}{\partial M_{\tilde t_2}^2}=\frac{3M_{\tilde t_2}^2}{16\pi^2}(\log\frac{M_{\tilde t_2}^2}{Q^2}-1),\;\;\frac{\partial^2 V^{(1)}}{\partial M_{\tilde t_2}^2\partial M_{\tilde t_2}^2}=\frac{3}{16\pi^2}\log\frac{M_{\tilde t_2}^2}{Q^2},
\end{eqnarray}
and the results of $\partial V^{(2)}/\partial M^2$, $\partial^2 V^{(2)}/(\partial M^2\partial M^2)$ with $M=m_t,\;M_{\tilde t_1},\;M_{\tilde t_2}$ can be found in the appendix of Ref.~\cite{Degrassi:2009yq}. The derivatives of $m_t^2,\;M_{\tilde t_1}^2,\;M_{\tilde t_2}^2,\;C_{2\bar\theta_t}^2$ over $\phi_i,\;(i=d,u,\eta,\bar\eta)$ can be written as
\begin{eqnarray}
&&\frac{\partial m_t^2}{\partial \phi_u}=Y_t^2 v_u,\;\;\frac{\partial^2 m_t^2}{\partial \phi_u\partial \phi_u}=Y_t^2,\\
&&\frac{\partial m^2_{\tilde t1,2}}{\partial \phi_d}=\frac{1}{8}G^2\phi_d\pm\frac{\frac{1}{24}G_1^2\phi_dM_{\tilde t LR}^2+Y_t \mu m_{\tilde t12}^2}{\Big\{M_{\tilde t LR}^4+2m_{\tilde t12}^4\Big\}^{1/2}},\nonumber\\
&&\frac{\partial^2 m^2_{\tilde t1,2}}{\partial \phi_d\partial \phi_d}=\frac{1}{8}G^2\mp\frac{2\Big\{-\frac{1}{24}G_1^2v_dM_{\tilde t LR}^2+Y_t \mu m_{\tilde t12}^2\Big\}^2}{\Big\{M_{\tilde t LR}^4+2m_{\tilde t12}^4\Big\}^{3/2}}\mp\frac{\frac{1}{24}G_1^2M_{\tilde t LR}^2-\frac{2}{24^2}G_1^4v_d^2-(\mu Y_t)^2}{\Big\{M_{\tilde t LR}^4+2m_{\tilde t12}^4\Big\}^{1/2}},\\
&&\frac{\partial m^2_{\tilde t1,2}}{\partial \phi_u}=(Y_t^2-\frac{1}{8}G^2)v_u\pm\frac{\frac{1}{24}G_1^2v_uM_{\tilde t LR}^2-Y_tA_t m_{\tilde t12}^2}{\Big\{M_{\tilde t LR}^4+2m_{\tilde t12}^4\Big\}^{1/2}},\nonumber\\
&&\frac{\partial^2 m^2_{\tilde t1,2}}{\partial \phi_u\partial \phi_u}=Y_t^2-\frac{1}{8}G^2\mp\frac{2\Big\{\frac{1}{24}G_1^2v_uM_{\tilde t LR}^2-Y_t A_t m_{\tilde t12}^2\Big\}^2}{\Big\{M_{\tilde t LR}^4+2m_{\tilde t12}^4\Big\}^{3/2}}\pm\frac{\frac{1}{24}G_1^2M_{\tilde t LR}^2+\frac{2}{24^2}G_1^4v_u^2+(A_t Y_t)^2}{\Big\{M_{\tilde t LR}^4+2m_{\tilde t12}^4\Big\}^{1/2}},\nonumber\\
&&\\
&&\frac{\partial m^2_{\tilde t1,2}}{\partial \phi_\eta}=\frac{1}{4}g_Bg_{YB}v_\eta\mp \frac{\frac{1}{24}G_2^2v_\eta M_{\tilde t LR}^2}{\Big\{M_{\tilde t LR}^4+2m_{\tilde t12}^4\Big\}^{1/2}},\nonumber\\
&&\frac{\partial^2 m^2_{\tilde t1,2}}{\partial \phi_\eta\partial \phi_\eta}=\frac{1}{4}g_Bg_{YB}\pm\frac{(\frac{1}{12}G_2^2m_{\tilde t12}^2v_\eta)^2}{\Big\{M_{\tilde t LR}^4+2m_{\tilde t12}^4\Big\}^{3/2}}\mp\frac{\frac{1}{24}G_2^2M_{\tilde t LR}^2}{\Big\{M_{\tilde t LR}^4+2m_{\tilde t12}^4\Big\}^{1/2}},\\
&&\frac{\partial m^2_{\tilde t1,2}}{\partial \phi_{\bar\eta}}=-\frac{1}{4}g_Bg_{YB}v_{\bar\eta}\pm\frac{\frac{1}{24}G_2^2v_{\bar\eta}M_{\tilde t LR}^2}{\Big\{M_{\tilde t LR}^4+2m_{\tilde t12}^4\Big\}^{1/2}},\nonumber\\
&&\frac{\partial^2 m^2_{\tilde t1,2}}{\partial \phi_{\bar\eta}\partial \phi_{\bar\eta}}=-\frac{1}{4}g_Bg_{YB}\pm\frac{(\frac{1}{12}G_2^2m_{\tilde t12}^2v_{\bar\eta})^2}{\Big\{M_{\tilde t LR}^4+2m_{\tilde t12}^4\Big\}^{3/2}}\pm\frac{\frac{1}{24}G_2^2M_{\tilde t LR}^2}{\Big\{M_{\tilde t LR}^4+2m_{\tilde t12}^4\Big\}^{1/2}},\\
&&\frac{\partial^2 m^2_{\tilde t1,2}}{\partial \phi_d\partial \phi_u}=\mp (Y_tA_tM_{\tilde t LR}^2+\frac{1}{12}G_1^2v_u m_{\tilde t12}^2)(Y_t \mu M_{\tilde t LR}^2+\frac{1}{12}G_1^2v_d m_{\tilde t12}^2)/(M_{\tilde t LR}^4+2m_{\tilde t12}^4)^{3/2},\nonumber\\
&&\\
&&\frac{\partial^2 m^2_{\tilde t1,2}}{\partial \phi_d\partial \phi_\eta}=\pm\frac{1}{12}G_2^2v_\eta m_{\tilde t12}^2\Big\{Y_t\mu M_{\tilde t LR}^2+\frac{1}{12}G_1^2v_dm_{\tilde t12}^2\Big\}/\Big\{M_{\tilde t LR}^4+2m_{\tilde t12}^4\Big\}^{3/2},\\
&&\frac{\partial^2 m^2_{\tilde t1,2}}{\partial \phi_d\partial \phi_{\bar\eta}}=\mp\frac{1}{12}G_2^2v_{\bar\eta} m_{\tilde t12}^2\Big\{Y_t\mu M_{\tilde t LR}^2+\frac{1}{12}G_1^2v_dm_{\tilde t12}^2\Big\}/\Big\{M_{\tilde t LR}^4+2m_{\tilde t12}^4\Big\}^{3/2},\\
&&\frac{\partial^2 m^2_{\tilde t1,2}}{\partial \phi_u\partial \phi_\eta}=\mp\frac{1}{12}G_2^2v_\eta m_{\tilde t12}^2\Big\{Y_tA_tM_{\tilde t LR}^2+\frac{1}{12}G_1^2v_um_{\tilde t12}^2\Big\}/\Big\{M_{\tilde t LR}^4+2m_{\tilde t12}^4\Big\}^{3/2},\\
&&\frac{\partial^2 m^2_{\tilde t1,2}}{\partial \phi_u\partial \phi_{\bar\eta}}=\pm\frac{1}{12}G_2^2v_{\bar\eta} m_{\tilde t12}^2\Big\{Y_tA_tM_{\tilde t LR}^2+\frac{1}{12}G_1^2v_um_{\tilde t12}^2\Big\}/\Big\{M_{\tilde t LR}^4+2m_{\tilde t12}^4\Big\}^{3/2},\\
&&\frac{\partial^2 m^2_{\tilde t1,2}}{\partial \phi_\eta\partial \phi_{\bar\eta}}=\mp\frac{(\frac{1}{12}G_2^2m_{\tilde t12}^2)^2v_\eta v_{\bar\eta}}{\Big\{M_{\tilde t LR}^4+2m_{\tilde t12}^4\Big\}^{3/2}},\\
&&\frac{\partial C_{2\bar\theta_t}^2}{\partial \phi_d}=\frac{4m_{\tilde t_{12}}^2}{(M_{\tilde t_1}^2-M_{\tilde t_2}^2)^3}\Big[m_{\tilde t_{12}}^2(\frac{\partial M_{\tilde t_1}^2}{\partial \phi_d}-\frac{\partial M_{\tilde t_2}^2}{\partial \phi_d})-(M_{\tilde t_1}^2-M_{\tilde t_2}^2)\frac{\partial m_{\tilde t_{12}}^2}{\partial \phi_d}\Big],\nonumber\\
&&\frac{\partial^2 C_{2\bar\theta_t}^2}{\partial \phi_d\partial \phi_d}=\frac{4}{(M_{\tilde t_1}^2-M_{\tilde t_2}^2)^4}\Big[(M_{\tilde t_1}^2-M_{\tilde t_2}^2)\frac{\partial m_{\tilde t_{12}}^2}{\partial \phi_d}-3m_{\tilde t_{12}}^2(\frac{\partial M_{\tilde t_1}^2}{\partial \phi_d}-\frac{\partial M_{\tilde t_2}^2}{\partial \phi_d})\Big]\Big[m_{\tilde t_{12}}^2(\frac{\partial M_{\tilde t_1}^2}{\partial \phi_d}\nonumber\\
&&\qquad\quad\;\;-\frac{\partial M_{\tilde t_2}^2}{\partial \phi_d})-(M_{\tilde t_1}^2-M_{\tilde t_2}^2)\frac{\partial m_{\tilde t_{12}}^2}{\partial \phi_d}\Big]+\frac{4m_{\tilde t_{12}}^2}{(M_{\tilde t_1}^2-M_{\tilde t_2}^2)^3}\Big[m_{\tilde t_{12}}^2(\frac{\partial^2 M_{\tilde t_1}^2}{\partial \phi_d\partial \phi_d}-\frac{\partial^2 M_{\tilde t_2}^2}{\partial \phi_d\partial \phi_d})\Big],\\
&&\frac{\partial C_{2\bar\theta_t}^2}{\partial \phi_u}=\frac{4m_{\tilde t_{12}}^2}{(M_{\tilde t_1}^2-M_{\tilde t_2}^2)^3}\Big[m_{\tilde t_{12}}^2(\frac{\partial M_{\tilde t_1}^2}{\partial \phi_u}-\frac{\partial M_{\tilde t_2}^2}{\partial \phi_u})-(M_{\tilde t_1}^2-M_{\tilde t_2}^2)\frac{\partial m_{\tilde t_{12}}^2}{\partial \phi_u}\Big],\nonumber\\
&&\frac{\partial^2 C_{2\bar\theta_t}^2}{\partial \phi_u\partial \phi_u}=\frac{4}{(M_{\tilde t_1}^2-M_{\tilde t_2}^2)^4}\Big[(M_{\tilde t_1}^2-M_{\tilde t_2}^2)\frac{\partial m_{\tilde t_{12}}^2}{\partial \phi_u}-3m_{\tilde t_{12}}^2(\frac{\partial M_{\tilde t_1}^2}{\partial \phi_u}-\frac{\partial M_{\tilde t_2}^2}{\partial \phi_u})\Big]\Big[m_{\tilde t_{12}}^2(\frac{\partial M_{\tilde t_1}^2}{\partial \phi_u}\nonumber\\
&&\qquad\quad\;\;-\frac{\partial M_{\tilde t_2}^2}{\partial \phi_u})-(M_{\tilde t_1}^2-M_{\tilde t_2}^2)\frac{\partial m_{\tilde t_{12}}^2}{\partial \phi_u}\Big]+\frac{4m_{\tilde t_{12}}^2}{(M_{\tilde t_1}^2-M_{\tilde t_2}^2)^3}\Big[m_{\tilde t_{12}}^2(\frac{\partial^2 M_{\tilde t_1}^2}{\partial \phi_u\partial \phi_u}-\frac{\partial^2 M_{\tilde t_2}^2}{\partial \phi_u\partial \phi_u})\Big],\\
&&\frac{\partial C_{2\bar\theta_t}^2}{\partial \phi_\eta}=\frac{4m_{\tilde t_{12}}^2}{(M_{\tilde t_1}^2-M_{\tilde t_2}^2)^3}\Big[m_{\tilde t_{12}}^2(\frac{\partial M_{\tilde t_1}^2}{\partial \phi_\eta}-\frac{\partial M_{\tilde t_2}^2}{\partial \phi_\eta})\Big],\nonumber\\
&&\frac{\partial^2 C_{2\bar\theta_t}^2}{\partial \phi_\eta\partial \phi_\eta}=\frac{4}{(M_{\tilde t_1}^2-M_{\tilde t_2}^2)^4}\Big[-3m_{\tilde t_{12}}^2(\frac{\partial M_{\tilde t_1}^2}{\partial \phi_\eta}-\frac{\partial M_{\tilde t_2}^2}{\partial \phi_\eta})\Big]\Big[m_{\tilde t_{12}}^2(\frac{\partial M_{\tilde t_1}^2}{\partial \phi_\eta}-\frac{\partial M_{\tilde t_2}^2}{\partial \phi_\eta})\Big]\nonumber\\
&&\qquad\quad\;\;+\frac{4m_{\tilde t_{12}}^2}{(M_{\tilde t_1}^2-M_{\tilde t_2}^2)^3}\Big[m_{\tilde t_{12}}^2(\frac{\partial^2 M_{\tilde t_1}^2}{\partial \phi_\eta\partial \phi_\eta}-\frac{\partial^2 M_{\tilde t_2}^2}{\partial \phi_\eta\partial \phi_\eta})\Big],\\
&&\frac{\partial C_{2\bar\theta_t}^2}{\partial \phi_{\bar\eta}}=\frac{4m_{\tilde t_{12}}^2}{(M_{\tilde t_1}^2-M_{\tilde t_2}^2)^3}\Big[m_{\tilde t_{12}}^2(\frac{\partial M_{\tilde t_1}^2}{\partial \phi_{\bar\eta}}-\frac{\partial M_{\tilde t_2}^2}{\partial \phi_{\bar\eta}})\Big],\nonumber\\
&&\frac{\partial^2 C_{2\bar\theta_t}^2}{\partial \phi_{\bar\eta}\partial \phi_{\bar\eta}}=\frac{4}{(M_{\tilde t_1}^2-M_{\tilde t_2}^2)^4}\Big[-3m_{\tilde t_{12}}^2(\frac{\partial M_{\tilde t_1}^2}{\partial \phi_{\bar\eta}}-\frac{\partial M_{\tilde t_2}^2}{\partial \phi_{\bar\eta}})\Big]\Big[m_{\tilde t_{12}}^2(\frac{\partial M_{\tilde t_1}^2}{\partial \phi_{\bar\eta}}-\frac{\partial M_{\tilde t_2}^2}{\partial \phi_{\bar\eta}})\Big]\nonumber\\
&&\qquad\quad\;\;+\frac{4m_{\tilde t_{12}}^2}{(M_{\tilde t_1}^2-M_{\tilde t_2}^2)^3}\Big[m_{\tilde t_{12}}^2(\frac{\partial^2 M_{\tilde t_1}^2}{\partial \phi_{\bar\eta}\partial \phi_{\bar\eta}}-\frac{\partial^2 M_{\tilde t_2}^2}{\partial \phi_{\bar\eta}\partial \phi_{\bar\eta}})\Big],\\
&&\frac{\partial^2 C_{2\bar\theta_t}^2}{\partial \phi_d\partial \phi_u}=\frac{4}{(M_{\tilde t_1}^2-M_{\tilde t_2}^2)^4}\Big[(M_{\tilde t_1}^2-M_{\tilde t_2}^2)\frac{\partial m_{\tilde t_{12}}^2}{\partial \phi_u}-3m_{\tilde t_{12}}^2(\frac{\partial M_{\tilde t_1}^2}{\partial \phi_u}-\frac{\partial M_{\tilde t_2}^2}{\partial \phi_u})\Big]\Big[m_{\tilde t_{12}}^2(\frac{\partial M_{\tilde t_1}^2}{\partial \phi_d}\nonumber\\
&&\qquad\quad\;\;-\frac{\partial M_{\tilde t_2}^2}{\partial \phi_d})-(M_{\tilde t_1}^2-M_{\tilde t_2}^2)\frac{\partial m_{\tilde t_{12}}^2}{\partial \phi_d}\Big]+\frac{4m_{\tilde t_{12}}^2}{(M_{\tilde t_1}^2-M_{\tilde t_2}^2)^3}\Big[m_{\tilde t_{12}}^2(\frac{\partial^2 M_{\tilde t_1}^2}{\partial \phi_d\partial \phi_u}-\frac{\partial^2 M_{\tilde t_2}^2}{\partial \phi_d\partial \phi_u})\nonumber\\
&&\qquad\quad\;\;+\frac{\partial m_{\tilde t_{12}}^2}{\partial \phi_u}(\frac{\partial M_{\tilde t_1}^2}{\partial \phi_d}-\frac{\partial M_{\tilde t_2}^2}{\partial \phi_d})-\frac{\partial m_{\tilde t_{12}}^2}{\partial \phi_d}(\frac{\partial M_{\tilde t_1}^2}{\partial \phi_u}-\frac{\partial M_{\tilde t_2}^2}{\partial \phi_u})\Big],\\
&&\frac{\partial^2 C_{2\bar\theta_t}^2}{\partial \phi_d\partial \phi_\eta}=\frac{4}{(M_{\tilde t_1}^2-M_{\tilde t_2}^2)^4}\Big[-3m_{\tilde t_{12}}^2(\frac{\partial M_{\tilde t_1}^2}{\partial \phi_\eta}-\frac{\partial M_{\tilde t_2}^2}{\partial \phi_\eta})\Big]\Big[m_{\tilde t_{12}}^2(\frac{\partial M_{\tilde t_1}^2}{\partial \phi_d}-\frac{\partial M_{\tilde t_2}^2}{\partial \phi_d})\nonumber\\
&&\qquad\quad\;\;-(M_{\tilde t_1}^2-M_{\tilde t_2}^2)\frac{\partial m_{\tilde t_{12}}^2}{\partial \phi_d}\Big]+\frac{4m_{\tilde t_{12}}^2}{(M_{\tilde t_1}^2-M_{\tilde t_2}^2)^3}\Big[m_{\tilde t_{12}}^2(\frac{\partial^2 M_{\tilde t_1}^2}{\partial \phi_d\partial \phi_\eta}-\frac{\partial^2 M_{\tilde t_2}^2}{\partial \phi_d\partial \phi_\eta})\nonumber\\
&&\qquad\quad\;\;-\frac{\partial m_{\tilde t_{12}}^2}{\partial \phi_d}(\frac{\partial M_{\tilde t_1}^2}{\partial \phi_\eta}-\frac{\partial M_{\tilde t_2}^2}{\partial \phi_\eta})\Big],\\
&&\frac{\partial^2 C_{2\bar\theta_t}^2}{\partial \phi_d\partial \phi_{\bar\eta}}=\frac{4}{(M_{\tilde t_1}^2-M_{\tilde t_2}^2)^4}\Big[-3m_{\tilde t_{12}}^2(\frac{\partial M_{\tilde t_1}^2}{\partial \phi_{\bar\eta}}-\frac{\partial M_{\tilde t_2}^2}{\partial \phi_{\bar\eta}})\Big]\Big[m_{\tilde t_{12}}^2(\frac{\partial M_{\tilde t_1}^2}{\partial \phi_d}-\frac{\partial M_{\tilde t_2}^2}{\partial \phi_d})\nonumber\\
&&\qquad\quad\;\;-(M_{\tilde t_1}^2-M_{\tilde t_2}^2)\frac{\partial m_{\tilde t_{12}}^2}{\partial \phi_d}\Big]+\frac{4m_{\tilde t_{12}}^2}{(M_{\tilde t_1}^2-M_{\tilde t_2}^2)^3}\Big[m_{\tilde t_{12}}^2(\frac{\partial^2 M_{\tilde t_1}^2}{\partial \phi_d\partial \phi_{\bar\eta}}-\frac{\partial^2 M_{\tilde t_2}^2}{\partial \phi_d\partial \phi_{\bar\eta}})\nonumber\\
&&\qquad\quad\;\;-\frac{\partial m_{\tilde t_{12}}^2}{\partial \phi_d}(\frac{\partial M_{\tilde t_1}^2}{\partial \phi_{\bar\eta}}-\frac{\partial M_{\tilde t_2}^2}{\partial \phi_{\bar\eta}})\Big],\\
&&\frac{\partial^2 C_{2\bar\theta_t}^2}{\partial \phi_u\partial \phi_\eta}=\frac{4}{(M_{\tilde t_1}^2-M_{\tilde t_2}^2)^4}\Big[-3m_{\tilde t_{12}}^2(\frac{\partial M_{\tilde t_1}^2}{\partial \phi_\eta}-\frac{\partial M_{\tilde t_2}^2}{\partial \phi_\eta})\Big]\Big[m_{\tilde t_{12}}^2(\frac{\partial M_{\tilde t_1}^2}{\partial \phi_u}-\frac{\partial M_{\tilde t_2}^2}{\partial \phi_u})\nonumber\\
&&\qquad\quad\;\;-(M_{\tilde t_1}^2-M_{\tilde t_2}^2)\frac{\partial m_{\tilde t_{12}}^2}{\partial \phi_u}\Big]+\frac{4m_{\tilde t_{12}}^2}{(M_{\tilde t_1}^2-M_{\tilde t_2}^2)^3}\Big[m_{\tilde t_{12}}^2(\frac{\partial^2 M_{\tilde t_1}^2}{\partial \phi_u\partial \phi_\eta}-\frac{\partial^2 M_{\tilde t_2}^2}{\partial \phi_u\partial \phi_\eta})\nonumber\\
&&\qquad\quad\;\;-\frac{\partial m_{\tilde t_{12}}^2}{\partial \phi_u}(\frac{\partial M_{\tilde t_1}^2}{\partial \phi_\eta}-\frac{\partial M_{\tilde t_2}^2}{\partial \phi_\eta})\Big],\\
&&\frac{\partial^2 C_{2\bar\theta_t}^2}{\partial \phi_u\partial \phi_{\bar\eta}}=\frac{4}{(M_{\tilde t_1}^2-M_{\tilde t_2}^2)^4}\Big[-3m_{\tilde t_{12}}^2(\frac{\partial M_{\tilde t_1}^2}{\partial \phi_{\bar\eta}}-\frac{\partial M_{\tilde t_2}^2}{\partial \phi_{\bar\eta}})\Big]\Big[m_{\tilde t_{12}}^2(\frac{\partial M_{\tilde t_1}^2}{\partial \phi_u}-\frac{\partial M_{\tilde t_2}^2}{\partial \phi_u})\nonumber\\
&&\qquad\quad\;\;-(M_{\tilde t_1}^2-M_{\tilde t_2}^2)\frac{\partial m_{\tilde t_{12}}^2}{\partial \phi_u}\Big]+\frac{4m_{\tilde t_{12}}^2}{(M_{\tilde t_1}^2-M_{\tilde t_2}^2)^3}\Big[m_{\tilde t_{12}}^2(\frac{\partial^2 M_{\tilde t_1}^2}{\partial \phi_u\partial \phi_{\bar\eta}}-\frac{\partial^2 M_{\tilde t_2}^2}{\partial \phi_u\partial \phi_{\bar\eta}})\nonumber\\
&&\qquad\quad\;\;-\frac{\partial m_{\tilde t_{12}}^2}{\partial \phi_u}(\frac{\partial M_{\tilde t_1}^2}{\partial \phi_{\bar\eta}}-\frac{\partial M_{\tilde t_2}^2}{\partial \phi_{\bar\eta}})\Big],\\
&&\frac{\partial^2 C_{2\bar\theta_t}^2}{\partial \phi_\eta\partial \phi_{\bar\eta}}=\frac{4}{(M_{\tilde t_1}^2-M_{\tilde t_2}^2)^4}\Big[-3m_{\tilde t_{12}}^2(\frac{\partial M_{\tilde t_1}^2}{\partial \phi_{\bar\eta}}-\frac{\partial M_{\tilde t_2}^2}{\partial \phi_{\bar\eta}})\Big]\Big[m_{\tilde t_{12}}^2(\frac{\partial M_{\tilde t_1}^2}{\partial \phi_\eta}-\frac{\partial M_{\tilde t_2}^2}{\partial \phi_\eta})\Big]\nonumber\\
&&\qquad\quad\;\;+\frac{4m_{\tilde t_{12}}^2}{(M_{\tilde t_1}^2-M_{\tilde t_2}^2)^3}\Big[m_{\tilde t_{12}}^2(\frac{\partial^2 M_{\tilde t_1}^2}{\partial \phi_\eta\partial \phi_{\bar\eta}}-\frac{\partial^2 M_{\tilde t_2}^2}{\partial \phi_\eta\partial \phi_{\bar\eta}})\Big],
\end{eqnarray}
where
\begin{eqnarray}
&&G^2=g_1^2+g_2^2+g_{YB}^2,\;G_1^2=5g_1^2-3g_2^2+5g_{YB}^2+2g_{B}g_{YB},\;G_2^2=4g_B^2+10g_Bg_{YB},\nonumber\\
&&m_{\tilde t 12}^2=Y_t(\mu v_d-A_t v_u),\nonumber\\
&&M_{\tilde t LR}^2=\frac{1}{24}[24(M_{Q}^2-M_{U}^2)+G_1^2(v_u^2-v_d^2)+G_2^2(v_{\bar\eta}^2-v_\eta^2)].
\end{eqnarray}

\section{The contributions from squarks to $S,\;T,\;U$ parameters in the B-LSSM.\label{A2}}

\begin{figure}
\setlength{\unitlength}{1mm}
\centering
\includegraphics[width=3in]{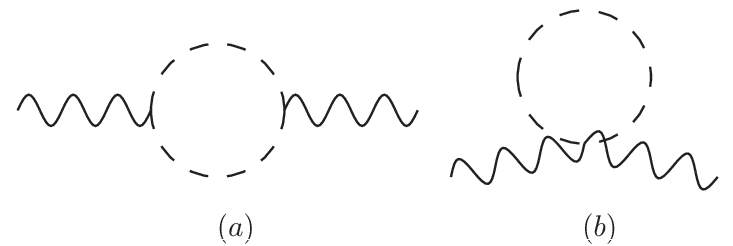}
\vspace{0cm}
\caption[]{The Feynman diagrams contributing to the $\gamma,\;W,\;Z$ self-energy and $\gamma-Z$ mixing, where the loop particles are squarks.}
\label{Wself}
\end{figure}

The one-loop diagrams contributing to the $\gamma,\;W,\;Z$ self-energy and $\gamma-Z$ mixing are plotted in Fig.~\ref{Wself}, where the loop particles are squarks.
Since the $Z-Z'$ mixing is limited to be very small ($\lesssim10^{-4}$) experimentally, we can neglect the $Z-Z'$ mixing contributions safely in the calculation. Then the results can be written as
\begin{eqnarray}
&&\Pi_{ZZ}=\frac{1}{16\pi^2}(C_{Z\tilde Q_i\tilde Q_j})^2\Big[I_1(M_{\tilde Q_i}^2,M_{\tilde Q_j}^2)+I_2(M_{\tilde Q_i}^2,M_{\tilde Q_j}^2)q^2\Big]-\nonumber\\
&&\qquad\quad\;\frac{1}{16\pi^2}(C_{ZZ\tilde Q_i\tilde Q_i})I_3(M_{\tilde Q_i}^2),\\
&&\Pi_{WW}=\frac{1}{16\pi^2}(C_{W\tilde U_i\tilde D_j})^2\Big[I_1(M_{\tilde U_i}^2,M_{\tilde D_j}^2)+I_2(M_{\tilde U_i}^2,M_{\tilde D_j}^2)q^2\Big]-\nonumber\\
&&\qquad\quad\;\;\frac{1}{16\pi^2}(C_{WW\tilde Q_i\tilde Q_i})I_3(M_{\tilde Q_i}^2),\\
&&\Pi_{\gamma\gamma}=\frac{1}{16\pi^2}(C_{\gamma\tilde Q_i\tilde Q_j})^2 I_2(M_{\tilde Q_i}^2,M_{\tilde Q_j}^2)q^2,\\
&&\Pi_{\gamma Z}=\frac{1}{16\pi^2}(C_{\gamma\tilde Q_i\tilde Q_j}C_{Z\tilde Q_i\tilde Q_j}) I_2(M_{\tilde Q_i}^2,M_{\tilde Q_j}^2)q^2,
\end{eqnarray}
where $C_{abc}$ and $C_{abcd}$ denote the coupling constants of fields $abc$, $abcd$ respectively, $M_{\tilde Q_i}$ denotes the squark mass, $M_{\tilde U_i}$ denotes the up-squark mass, $M_{\tilde D_i}$ denotes the down-squark mass, and
\begin{eqnarray}
&&I_1(x,y)=x({\rm div}-\ln x)+y({\rm div}-\ln y)+f(x,y),\\
&&I_2(x,y)=\frac{1}{18(x-y)^2}\Big[(x+y)^2+8xy+6x^2\ln x+6y^2\ln y\nonumber\\
&&\qquad\qquad\;-\frac{12xy}{x-y}(x\ln x-y \ln y)\Big],\\
&&I_3(x)=x({\rm div}-\ln x),
\end{eqnarray}
where
\begin{eqnarray}
&&{\rm div}\equiv\frac{2}{4-d}-\gamma+1+\ln(4\pi).
\end{eqnarray}
Then the $S,\;T,\;U$ parameters~\cite{Peskin:1991sw,Yang:2022gvz} at one-loop level in the B-LSSM can be written as
\begin{eqnarray}
&&\hat S=-\frac{c_W}{s_W}(c_W^2\Pi'_{\gamma Z}-s_W^2\Pi'_{\gamma Z}+c_Ws_W\Pi'_{\gamma\gamma}-
c_Ws_W\Pi'_{ZZ}),\nonumber\\
&&\hat T=\frac{\Pi_{33}-\Pi_{WW}}{M_W^2}=-\frac{1}{M_W^2}(c_W^2\Pi_{ZZ}+s_W^2\Pi_{\gamma\gamma}+2s_Wc_W\Pi_{\gamma Z}-
\Pi_{WW}),\nonumber\\
&&\hat U=c_W^2\Pi'_{ZZ}+s_W^2\Pi'_{\gamma\gamma}+2s_Wc_W\Pi'_{\gamma Z}-\Pi'_{WW},
\end{eqnarray}
where $\Pi_{V_1V_2}'$ denotes $\partial \Pi_{V_1V_2}/\partial q^2$ with $V_1V_2$ denoting $\gamma\gamma,\;ZZ,\;WW,\;\gamma Z$.

\begin{acknowledgments}
The work has been supported by the National Natural Science Foundation of China
(NNSFC) with Grants No. 12235008, No. 12675001, Natural Science Foundation for
Distinguished Young Scholars of Hebei Province with Grant No. A2023201041.
\end{acknowledgments}

\end{CJK*}

\end{document}